\documentclass[useAMS,usenatbib]{mn2e}
\usepackage{graphicx}

\rightline{\bf Version 3.3}

\title[Jovian Trojan Asteroids in SDSS MOC 3]
{The Properties of Jovian Trojan Asteroids Listed in \\ SDSS Moving Object Catalog 3}

\author[Gy. M. Szab\'o et al.]{Gy. M. Szab\'o\
$^{1}$\thanks{E-mail: szgy@titan.physx.u-szeged.hu},
\v{Z}. Ivezi\'c$^{2}$, M. Juri\'c$^{3}$, R. Lupton$^{3}$\\
$^{1}$Department of Experimental Physics \& Astronomical Observatory,
      University of Szeged,6720 Szeged, Hungary\\ 
      Magyary Zolt\'an Postdoctoral Research Fellow\\
$^{2}$Department of Astronomy, University of Washington, Seattle, WA 98155, USA\\
$^{3}$Princeton University Observatory, Princeton, NJ 08544, USA}

\begin{document}

\date{}
\pagerange{\pageref{firstpage}--\pageref{lastpage}} \pubyear{2002}
\maketitle
\begin{abstract}
We analyze 1187 observations of about 860 unique candidate Jovian Trojan
asteroids listed in the 3rd release of Sloan Digital Sky Survey (SDSS) Moving Object Catalog. 
The sample is complete at the faint end to $r=21.2$ mag (apparent brightness) 
and $H=13.8$ (absolute brightness, approximately corresponding to 10 km 
diameter). A subset of 297 detections of previously known Trojans were used 
to design and optimize a selection method based on observed angular velocity 
that resulted in the remaining objects. Using a sample of objects with known
orbits, we estimate that the candidate sample contamination is about 3\%. 
The well-controlled selection effects, the sample size, 
depth and accurate five-band UV-IR photometry enabled several new findings 
and the placement of older results on a firmer statistical footing. We find 
that there are significantly more asteroids in the leading swarm (L4) than
in the trailing swarm (L5): $N(L4)/N(L5)=1.6\pm0.1$, independently of 
limiting object's size. The overall counts normalization suggests that there are about
as many Jovians Trojans as there are main-belt asteroids down to the same 
size limit, in agreement with earlier estimates. We find that Trojan asteroids 
have a remarkably narrow color distribution (root-mean-scatter of only 
$\sim$0.05 mag) that is significantly different from the color distribution 
of the main-belt asteroids. The color of Trojan asteroids is correlated with 
their orbital inclination, in a similar way for both swarms, but appears 
uncorrelated with the object's size. We extrapolate the results presented
here and estimate that Large Synoptic Survey Telescope will determine orbits, 
accurate colors and measure light curves in six photometric bandpasses for 
about 100,000 Jovian Trojan asteroids. 
\end{abstract}



\section{Introduction.}

Jovian Trojan asteroids are found in two swarms around the L4 and L5 Lagrangian 
points of the Jupiter's orbit (for a review see Marzari et 
al. 2001). The first Jovian Trojan was discovered a century ago by Max Wolf. 
Close to 2,000 Jovian Trojans were discovered by the end of 2003 (Bendjoya et al. 
2004, hereafter B04). About half are numbered asteroids with reliable orbits 
(Marzari et al. 2001). Their total number is suspected to be similar to the number 
of the main belt asteroids\footnote{Recent work supports this claim. Ivezi\'{c} et al. 
(2001) estimated that the number of main-belt asteroids with diameters larger than 
1 km is 740,000, with a somewhat higher estimate by Tedesco, Cellino \& Zappal\'{a} (2005), 
and Jewitt, Trujillo \& Luu (2000) estimated that there are between 520,000 and 
790,000 Jovian Trojans above the same size limit.} (Shoemaker et al. 1989). 

Trojans' positions relative to Jupiter librate around L4 (leading swarm) and 
L5 (trailing swarm) with periods of the order hundred years. Their orbital 
eccentricity is typically smaller ($<$0.2) than those of main-belt asteroids, but 
the inclinations are comparable, with a few known Trojans having inclinations 
larger than 30 degree. The largest objects have diameters exceeding 100 km. 
They typically have featureless (D type) spectra and extremely low optical albedo 
(Tedesco 1989; Fern\'{a}ndez, Sheppard \& Jewitt 2003). These spectral
properties are similar to those of cometary nuclei. However, there are also
Trojans that have P or common C-type classification, mostly found in the 
trailing swarm (Fitzsimmons et al. 1994). The collisional grinding of Trojan 
asteroids is supported by their observed size distribution (Jewitt, Trujillo
\& Luu 2000, hereafter JTL). 

Numerous studies of the origin of Jovian Trojans are based on two different
hypothesis. According to one of them, the Jovian Trojans were formed
simultaneously with Jupiter in the early phase of the solar nebula. The growing 
Jupiter could have captured and stabilized the plantesimals near 
its L4 and L5 points (Peale 1993). The other hypothesis assumes that 
the majority of Jovian Trojans were captured over a much longer period,
and were formed either close to Jupiter, or were gravitationally 
scattered from the main belt or elsewhere in the solar system (Jewitt 1996). 
The spectral comet-like appearance of many Trojans is consistent with the 
scattering from the outer Solar System.

Depending on the importance of gas drag when Trojans formed, the L4 and L5 
swarms could have different dynamics. 
The presence 
of significant gas drag helps stabilize orbits around the L5 point.
On the other hand, these trailing objects have later evolution different 
from the leading swarm because planetary migration destabilizes L5 (Gomes 1998).
Morbidelli et al. (2005) recently suggested a more complex picture:
the present permanent Trojan populations are built up by objects that were
trapped after the 1:2 mean motion resonance crossing of the Saturn and the
Jupiter. Therefore, it is possible that size distributions, or detailed 
distributions of orbital parameters, could be different for the leading
and trailing swarm. However, no such differences have yet been
found (Marzari et al., 2001, and references therein).

It is noteworthy
that there are severe observational biases in the sample of known Jovian
Trojans due to their large distance. For example, although the numbers
of main-belt asteroids and Trojans to a given size limit are similar,
only about 1\% of the known objects belong to the latter group. This
is a consequence of the fact that a Trojan at a heliocentric distance of 
5.2 AU is about 4 magnitudes fainter than a same-size main-belt asteroid 
at a heliocentric distance of 2.5 AU (as observed in opposition, and not 
accounting for differences in albedo, which further diminishes the 
Trojan's apparent magnitude).  

Here we present an analysis of the properties of about 1000 known and 
candidate Jovian Trojan asteroids based on the data collected by Sloan 
Digital Sky Survey (SDSS, York et al. 2000). SDSS, although primarily 
designed for observations of extragalactic objects, is significantly 
contributing to studies of the solar system objects because asteroids 
in the imaging survey must be explicitly detected and measured to avoid 
contamination of the samples of extragalactic objects selected for
spectroscopy. Preliminary analysis of SDSS commissioning data (Ivezi\'{c} 
et al. 2001, hereafter I01) showed that SDSS will increase the number of
asteroids with accurate five-color photometry by more than two orders 
of magnitude, and to a limit about five magnitudes fainter (seven magnitudes
when the completeness limits are compared) than previous multi-color surveys
(e.g. The Eight Color Asteroid Survey, Zellner, Tholen \& Tedesco 1985).
As we demonstrate below, the SDSS data extend the faint completeness 
limit for Trojan asteroids by about 1.5 magnitudes (to a limiting diameter
of $\sim10$ km).  

The large sample and accurate astrometric and five-band photometric
SDSS data to a much fainter limit than reached by most previous surveys, 
together with suitable ways to quantify selection effects, allow
us to address the following questions:
\begin{enumerate}
\item What is the size distribution of Jovian Trojans asteroids with
      diameters larger than 10 km?
\item Do the leading and trailing swarms have the same size distribution
      (including both the distribution shape and the overall number above 
       some size limit)?
\item What is their color distribution in the SDSS photometric system, and how
      does it compare to the color distribution of main-belt asteroids?
\item Is the color distribution correlated with inclination, as suggested
      by a preliminary analysis of SDSS data (Ivezi\'{c} et al. 2002a, 
      hereafter I02a)? 
\item Are the Trojans' size and color correlated (as suggested by Bendjoya et al. 2004)?
\item Do the leading and trailing swarms have the same color distribution?
\item Is the size distribution correlated with inclination? 

\end{enumerate}

The SDSS asteroid data are described in Section 2, and in Section 3
we describe a novel method for selecting candidate Jovian Trojan asteroids 
from SDSS database. Analysis of the properties of selected objects,
guided by the above questions, is presented in Section 4. We summarize 
our results in Section 5, and discuss their implications for the origin 
and evolution of Trojan asteroids.

\section{           SDSS Observations of Moving Objects      }

\begin{figure} 
\centering
\includegraphics[width=8cm]{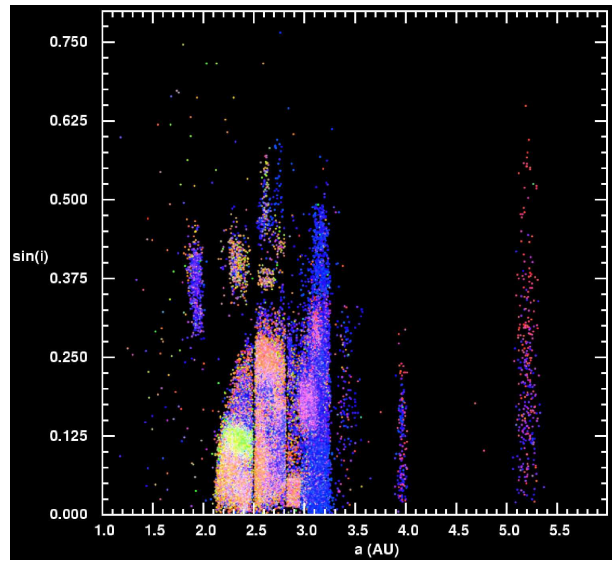}
\caption{
The dots show the osculating orbital inclination vs. semi-major axis distribution of 
43,424 unique moving objects detected by the SDSS, and matched to objects 
with known orbital parameters listed in Bowell's ASTORB file (these data
are publicly available in the third release of the SDSS Moving Object Catalog. 
The dots are color-coded according to their colors measured by SDSS (see I02a for 
details, including analogous figures constructed with proper orbital elements). 
Note that most main-belt asteroid families have distinctive colors. 
Jovian Trojans asteroids are found at $a$$\sim$5.2 AU, and display a 
correlation between the color and orbital inclination (objects with
high inclination tend to be redder, see Section 4).}
\label{pP1}
\end{figure}

\begin{figure} 
\centering
\includegraphics[width=8cm]{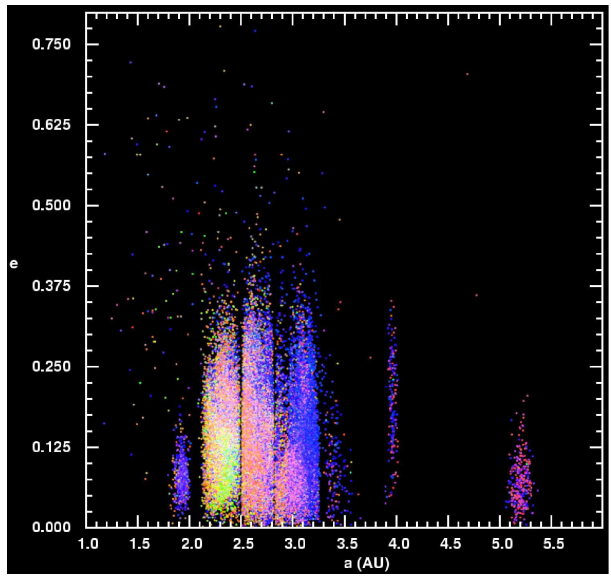}
\caption{
Analogous to Figure~\ref{pP1}, except that here the orbital  
eccentricity vs. semi-major axis distribution is shown. Note that there
is no discernible correlation between the color and eccentricity for 
Jovian Trojan asteroids.}
\label{prettyPlot2}
\end{figure}

SDSS is a digital photometric and spectroscopic survey using a dedicated 
2.5 m telescope at the Apache Point Observatory, which will cover
10,000 deg$^2$ of the Celestial Sphere in the North Galactic cap, and a
smaller ($\sim$ 225 deg$^2$) and deeper survey in the Southern Galactic
hemisphere (Abazajian et al. 2003, and references therein). The survey sky 
coverage will result in photometric measurements for over $10^8$ stars and 
a similar number of galaxies. The flux densities of detected objects are 
measured almost simultaneously (within $\sim$ five minutes) in five bands 
($u$, $g$, $r$, $i$, and $z$) with effective wavelengths 
of 3551 \AA, 4686 \AA, 6166 \AA, 7480 \AA, and 8932 \AA\ (Fukugita et al. 1996; 
Gunn et al.~1998; Smith et al.~2002; Hogg et al.~2002). The photometric 
catalogs are 95\% complete for point sources to limiting magnitudes of
22.0, 22.2, 22.2, 21.3, and 20.5 in the North Galactic cap. Astrometric
positions are accurate to about 0.1 arcsec per coordinate (rms) for sources
brighter than 20.5$^m$ (Pier et al. 2003), and the morphological information
from the images allows robust star-galaxy separation (Lupton et al. 2001, 2002) 
to $\sim$ 21.5$^m$. The photometric measurements are accurate to $0.02$ 
magnitudes (both absolute calibration, and root-mean-square scatter for 
sources not limited by photon statistics; Ivezi\'{c} et al.~2004).
The recent fifth public Data Release (DR5) includes imaging data for 
$\sim$8000 deg$^2$ of sky, and catalogs for $2.15 \times 10^8$ objects. For 
more details please see Abazajian et al. (2003) and references therein.

SDSS Moving Object Catalog\footnote{Available at http://www.sdss.org} (hereafter 
SDSS MOC) is a public, value-added catalog of SDSS asteroid observations
(Ivezi\'{c} et al. 2002b, hereafter I02b).
It includes all unresolved objects brighter than $r=21.5$ and with observed angular 
velocity in the 0.05--0.5 deg/day interval. In addition to providing SDSS astrometric 
and photometric measurements, all observations are matched to known objects listed 
in the ASTORB file (Bowell 2001), and to a database of proper orbital elements 
(Milani, 1999), as described in detail by Juri\'{c} et al. (2002, hereafter J02). 
J02 determined that the catalog completeness (number of moving objects 
detected by the software that are included in the catalog, divided by the total 
number of moving objects recorded in the images) is about 95\%, and its contamination 
rate is about 6\% (the number of entries that are not moving objects, but rather 
instrumental artifacts). 

The third release of SDSS MOC used in this work contains measurements for over 204,000 asteroids. 
The quality of these data was discussed in detail by I01, including a determination 
of the size and color distributions for main-belt asteroids. An analysis of 
correlation between colors and asteroid dynamical families was presented by 
I02a. An interpretation of this correlation as the dependence of color on family
age (due to space weathering effect) was proposed by Jedicke et al. (2004) and 
further discussed by Nesvorny et al. (2005). Multiple SDSS observations of objects 
with known orbital parameters can be accurately linked, and thus SDSS MOC also 
contains rich information about asteroid color variability, discussed in detail 
by Szab\'o et al. (2004). 

The value of SDSS data becomes particularly evident when exploring the
correlation between colors and orbital parameters for main-belt asteroids. 
Figure~\ref{pP1} uses a technique developed by I02a
to visualize this correlation. A striking feature of this
figure is the color homogeneity and distinctiveness displayed by asteroid 
families. This strong color segregation provides firm support for the 
reality of asteroid dynamical families. 
Jovian Trojans asteroids are found at $a$$\sim$5.2 AU, and display a 
correlation between the color and orbital inclination (objects with
high inclination tend to be redder). On the other hand, the color and 
orbital eccentricity (see Figure~\ref{prettyPlot2}) do not appear
correlated.

\begin{figure}
\centering
\includegraphics[width=8cm]{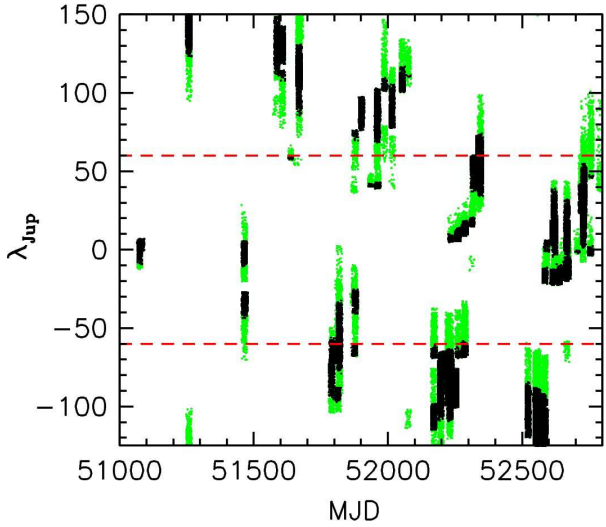}
\caption{The distribution of the longitude of $\sim$440,000 9$\times$13 
arcmin$^2$ large SDSS observing fields in a coordinate system center on Jupiter 
and aligned with its orbit, as a function of observing epoch (green symbols). 
Fields obtained within 25 deg. from the opposition are marked by black
symbols. The two dashed 
lines mark the relative longitudes of the L4 ($\lambda_{Jup}=60$ deg, leading
swarm) and L5 ($\lambda_{Jup}=-60$ deg, trailing swarm) Lagrangian points.
Both swarms are well sampled in the third release of SDSS Moving Object 
Catalog. 
\label{Lcoverage}}
\end{figure}

The distribution of the positions of SDSS observing fields in a coordinate 
system centered on Jupiter and aligned with its orbit is shown in Figure~\ref{Lcoverage}. 
As evident, both L4 and L5 regions are well covered with the available SDSS data. 
There are 313 unique known objects (from ASTORB file) in SDSS MOC whose orbital 
parameters are consistent with Jovian Trojan asteroids (here defined as objects 
with semi-major axis in the range 5.0--5.4 AU). Since SDSS imaging depth is about 
two magnitudes deeper than 
the completeness limit of ASTORB file used to identify known Trojans, there are 
many more Trojan asteroids in SDSS MOC whose orbits are presently unconstrained. 
Nevertheless, they can be identified using a kinematic method described in the 
following Section.

\section { Selection of Trojan asteroids from SDSS Moving Object Catalog }

\begin{figure} 
\centering
\includegraphics[width=8cm]{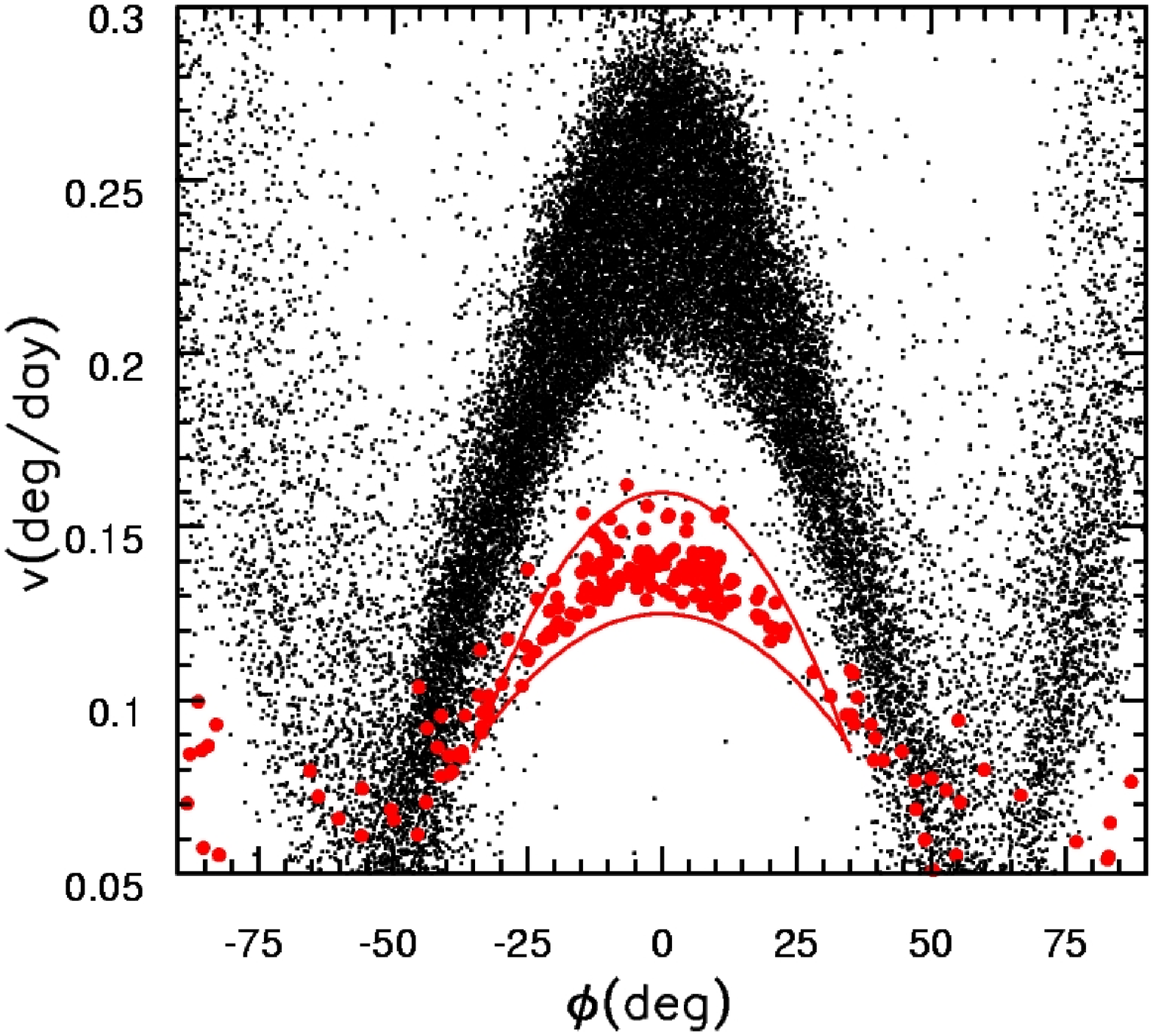}
\includegraphics[width=8cm]{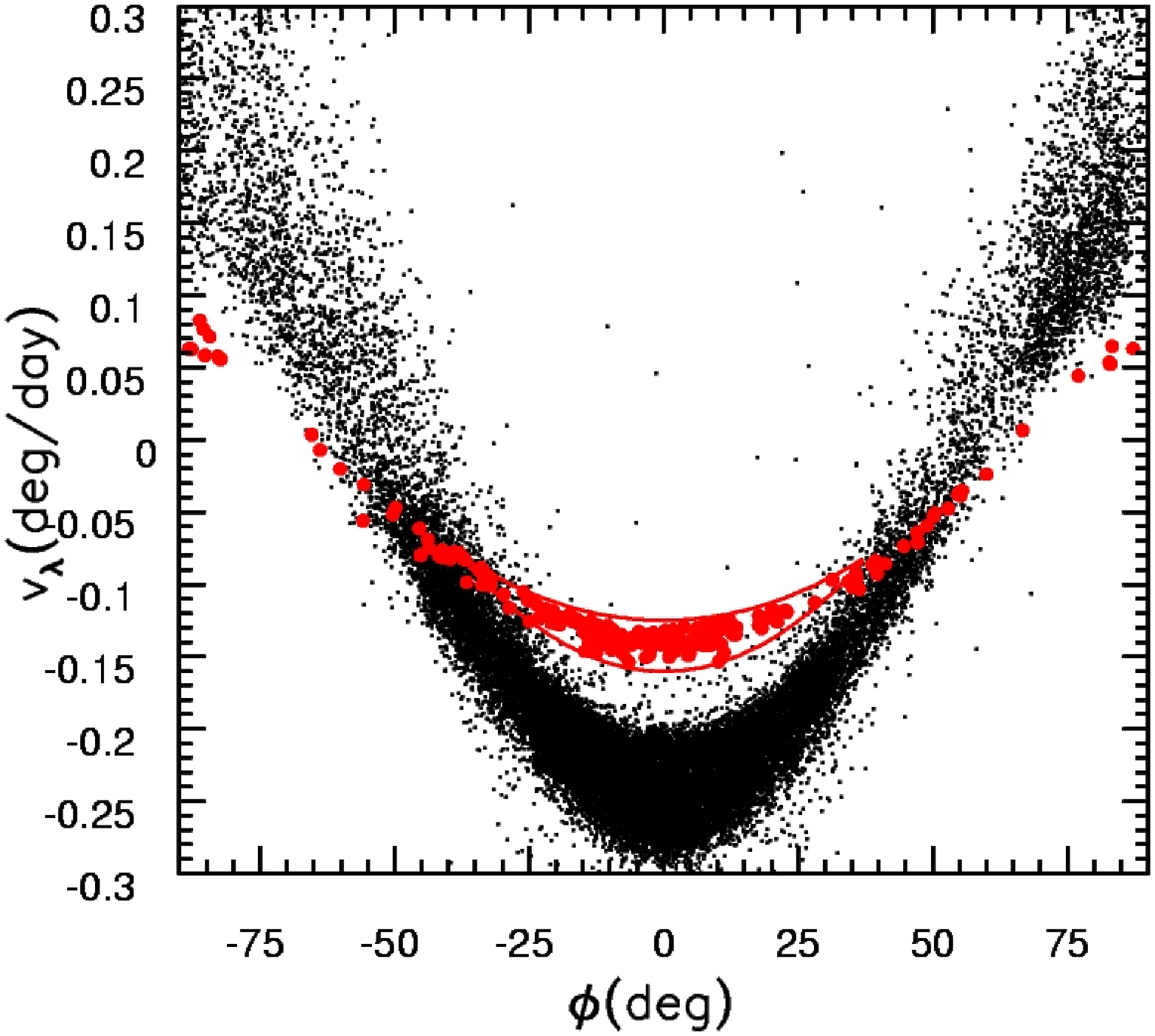}
\caption{The basis for the kinematic selection of candidate Trojan
asteroids from SDSS Moving Object Catalog. The small dots in the top
panel show the magnitude of the measured angular velocity as a function 
of the longitudinal angle from the opposition for $\sim$43,000 unique objects 
with known orbits listed in the catalog. The large dots show known Jovian 
Trojan asteroids. The lines show adopted selection criteria for candidate
Trojans (see text). The bottom panel is an analogous plot and shows the 
measured longitudinal component of the angular velocity (in ecliptic 
coordinate system) as a function of angle from the opposition. The 
candidate Trojans are selected in the three-dimensional $v-v_\lambda-\phi$ 
space.}
\label{knownKin}
\end{figure}

\begin{figure} 
\centering
\includegraphics[width=8cm]{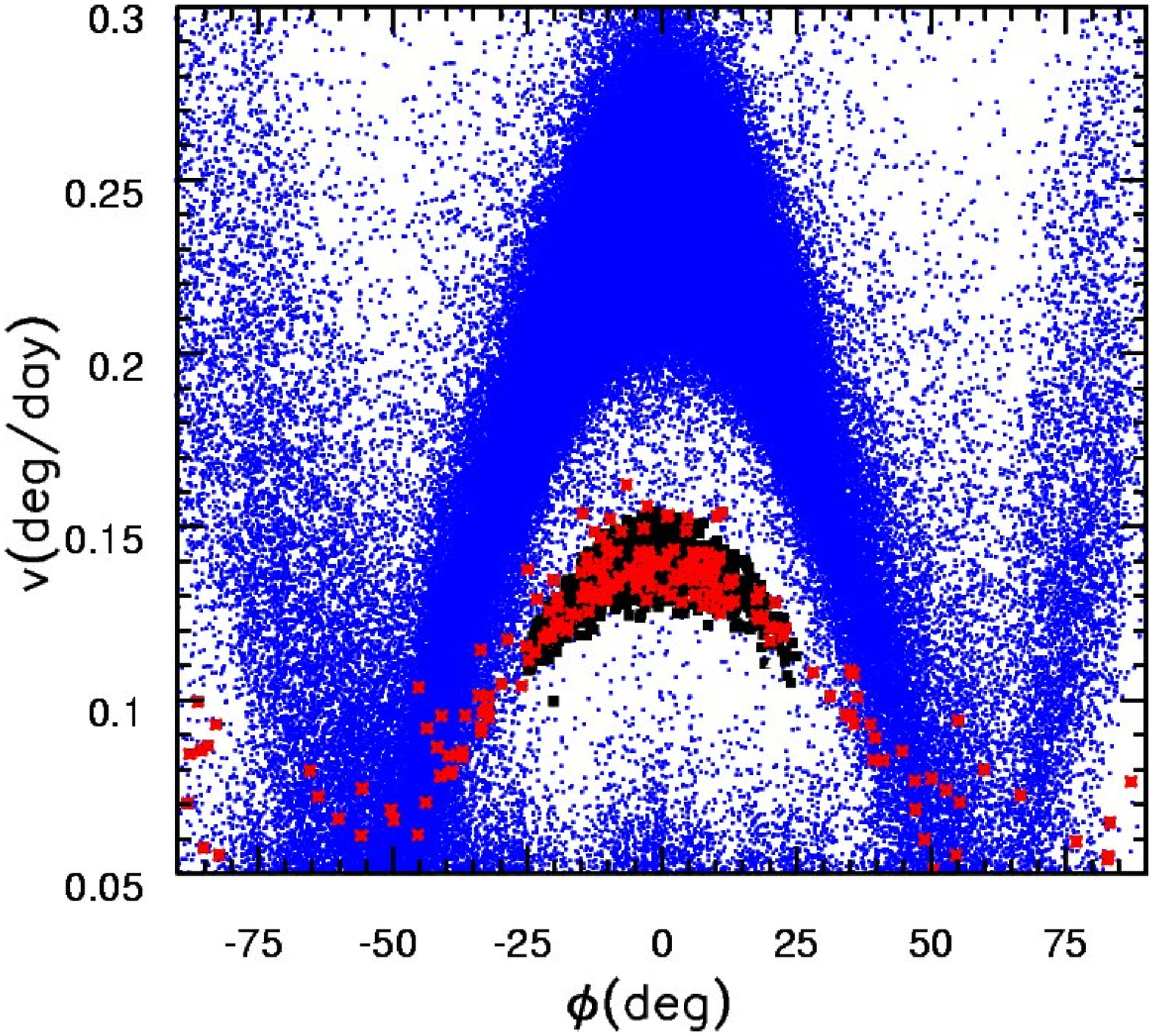}
\includegraphics[width=8cm]{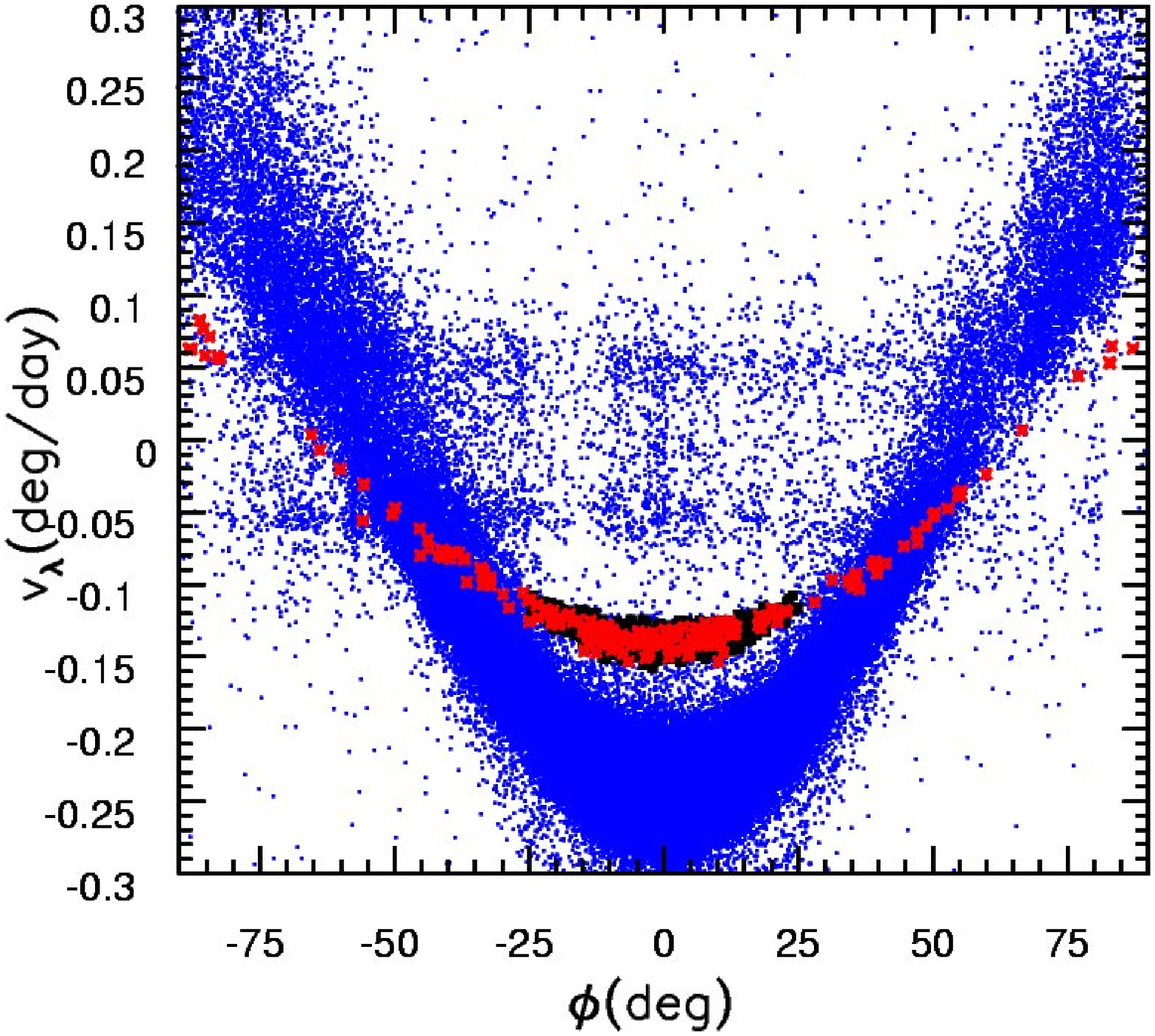}
\caption{Analogous to Figure~\ref{knownKin}, except that all $\sim$204,000
objects from SDSS Moving Object Catalog are shown (blue dots). The candidate
Trojans are shown by black symbols, and the known Trojans are overplotted
as red symbols.}
\label{allKin}
\end{figure}

The angular velocity of moving objects measured by SDSS can be used as a
proxy for their distance determination and classification (see Figure 14 
and Appendix A in I01). For example, Jovian Trojan
asteroids are typically slower than main-belt asteroids because their 
distances from Earth are larger (the observed angular velocity is dominated
by the Earth's reflex motion). However, in addition to angular 
velocity, the selection algorithm must also include the longitudinal angle 
from the opposition, $\phi$, because for large values of $|\phi|$ the
main-belt asteroids can have angular velocity as small as Jovian Trojans. 
This behavior is illustrated in Figure~\ref{knownKin}.

We optimize criteria for selecting candidate Jovian Trojans with the 
aid of 482 observations of 313 Trojans from SDSS MOC that have known 
orbits extracted from ASTORB file (there are 43,424 unique objects with 
known orbits in the third release of SDSS 
MOC). These 482 observations are identified in orbital space using 
constraints 5.0 AU $<$ a $<$ 5.4 AU and e $<$ 0.2, and hereafter
referred to as the Known Trojans (KT). Of those, the majority 
(263) belong to the leading swarm.

We compare the angular velocity and $\phi$ distributions of these objects 
to those for the whole sample in Figure~\ref{knownKin}. We find that 
the following selection criteria result in a good compromise between
the selection completeness and contamination:
\begin{eqnarray}
   0.112-{\left( \phi \over 180 \right) }^2 < {\rm v} < 0.155-{\left( 
       \phi \over 128 \right) }^2,  \\
   -0.160+{\left( \phi \over 134 \right) }^2 < {\rm v}_{\lambda} < - 
0.125+{\left( \phi \over 180 \right) }^2, 
\end{eqnarray}
for observations with $ -25 < \phi < 25.$. That is, only observations 
obtained relatively close to the opposition
can be used to select a sample with a low contamination rate by 
main-belt asteroids. The adopted velocity limits are in good agreement 
with those proposed by JTL.

When applied to all objects from SDSS MOC, this selection results in 
a sample of 1187 candidate Trojans, including 272 observations of known 
objects (see Figure~\ref{allKin}). Of the latter, 8 objects have semi-major 
axis too small to be a Trojan asteroid, which implies a contamination rate
of 3\%. SDSS MOC contains 297 observations of known Trojans obtained with 
$|\phi|<25$, which implies that the kinematic selection method is 89\% complete. 
The 264 detections of known Trojans in the kinematically selected sample
correspond to 191 unique objects. Therefore, 1187 detections in the candidate 
sample correspond to about 858 unique objects. 

The contamination rate could be higher than 3\% because objects with known 
orbits tend to be brighter and thus have smaller measurement uncertainties 
for angular velocities than objects from the full candidate sample (for a 
detailed study of these errors and their correlation with other observables 
see I01). For this reason, we perform the following robustness test. 

\begin{figure} 
\centering
\includegraphics[width=8cm]{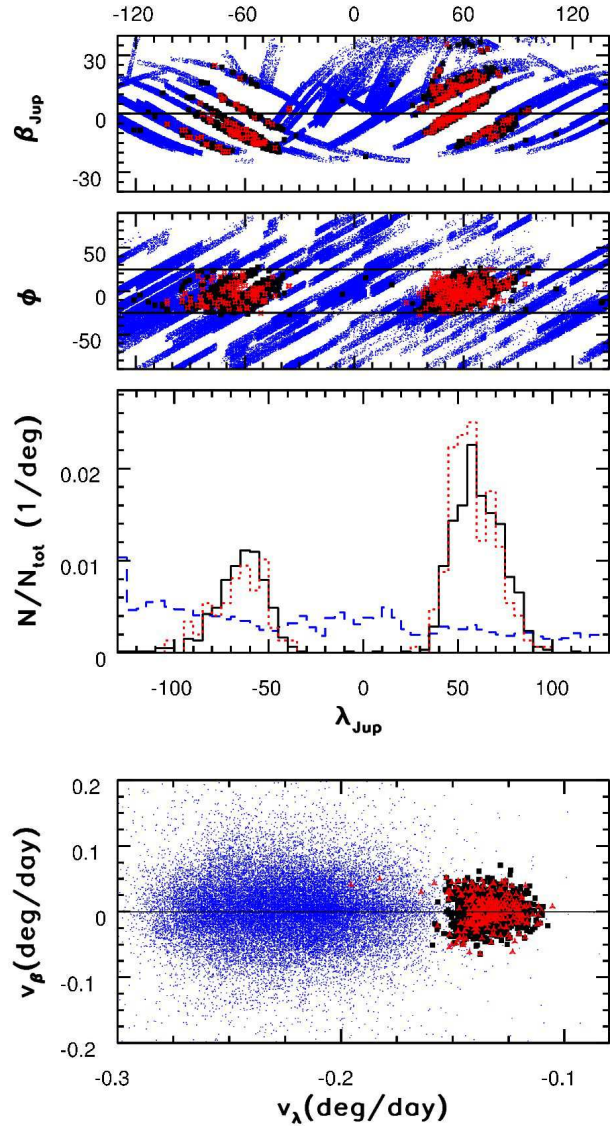}
\caption{A test of the selection robustness. The top two panels
show all the objects from SDSS MOC (small blue dots), the known Trojans 
(red dots) and the candidate Trojans (black squares), as observed on the 
sky, in Jupiter's coordinate system and in $\phi$ vs. $\lambda_{Jup}$ diagram. 
Although $\lambda_{Jup}$ was not used in selection, the known
and candidate Trojans have similar $\lambda_{Jup}$ distributions (the third
panel from top, dotted and solid histograms, respectively), and different
than for the whole sample, dominated by main-belt asteroids (dashed line). 
Note that these $\lambda_{Jup}$ distributions are not corrected for the
selection biases due to inhomogeneous coverage of $\lambda_{Jup}-\beta_{Jup}$
plane (which are presumably similar for both known and candidate
objects), and thus are not representative of the true distribution. 
The bottom panel compares the angular velocity distributions of Trojans
and main-belt asteroids.}
\label{lambdaTest}
\end{figure}

The above selection procedure does not include $\lambda_{Jup}$, 
the longitudinal angle between an object and Jupiter.
If the selection is robust, the $\lambda_{Jup}$ distributions for 
the known and candidate Trojans should be similar. As discernible 
from Figure~\ref{lambdaTest}, this is indeed the case and demonstrates that 
the contamination rate by non-Trojan asteroids in the candidate sample 
must be small. A similar conclusion is reached when comparing color
distributions (see below). We refer to this sample hereafter as the Candidate 
Trojans (CT). 

\section{     Analysis of the Properties of Trojan Asteroids       }

Using the sample of candidate Trojan asteroids selected as described 
above, here we analyze their distribution in the 3-dimensional 
size-color-inclination space, both for the full sample and separately for 
each swarm. The large size of the selected candidate sample allows 
accurate measurements of this distribution, and represents an especially 
significant improvement over the previous work when studying color distribution. 
The two largest homogeneous studies of spectral properties of Jovian Trojans are by 
Jewitt \& Luu (1990) and Bendjoya et al. (2004). Jewitt \& Luu
obtained spectra of 32 Trojans and found that they are remarkably similar 
to cometary spectra. Bendjoya et al.  obtained spectroscopic 
observations for 34 objects and, together with older observations, produced 
a sample of 73 objects. Therefore, accurate color information for over 
a thousand objects discussed here represents a substantial improvement.

\begin{figure*} 
\centering
\includegraphics[width=17.5cm]{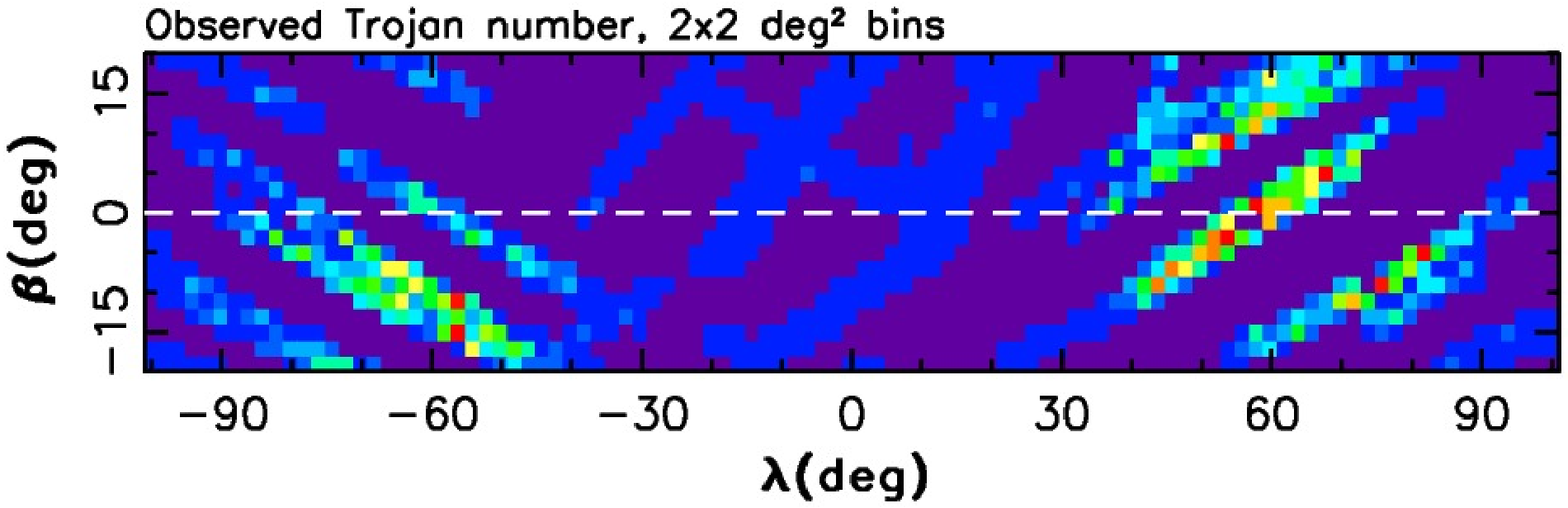}
\includegraphics[width=17.5cm]{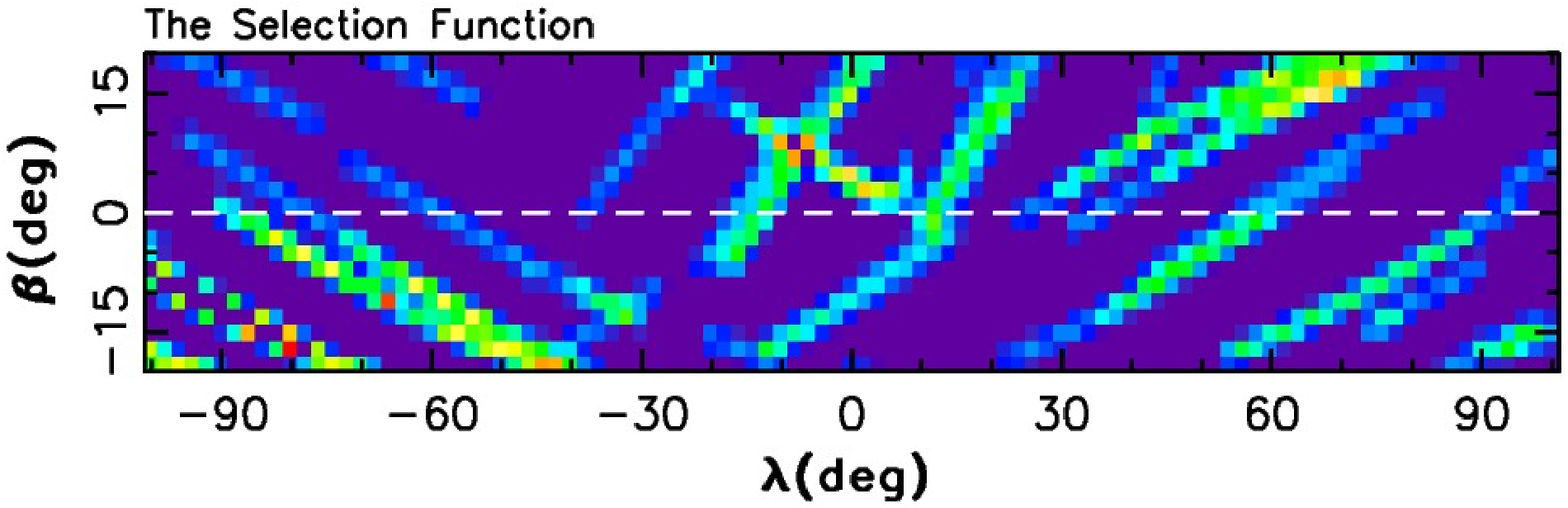}
\includegraphics[width=17.5cm]{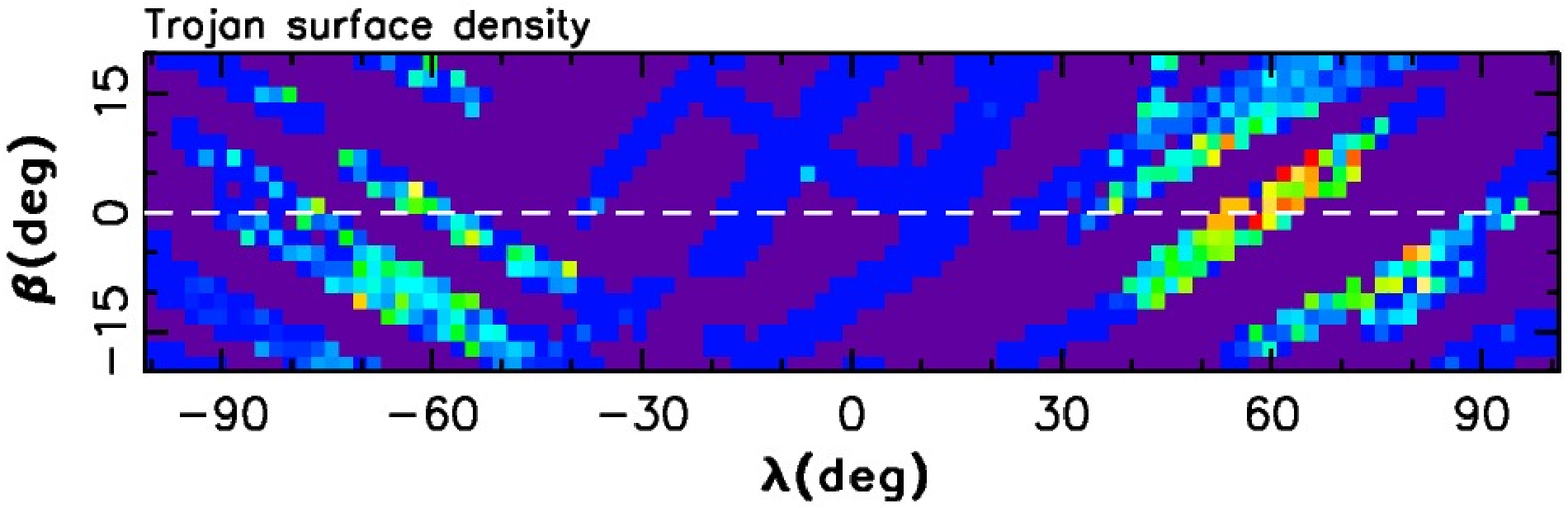}
\caption{The top panel shows the observed surface density map 
(number of detected objects per 4 deg$^2$ large bin) of candidate Trojan asteroids
in Jupiter's coordinate system. The middle panel shows the number of 
SDSS fields observed in each bin (that is, the selection function), 
and the bottom panel shows the corrected surface density of Trojans
(the ratio of the maps in the top and middle panels). The values
are shown on a linear scale, increasing from blue to red (i.e. no 
objects are found in blue strips). The maximum value (coded red) in the 
top panel is 20 (Trojans per 4 deg$^2$ large bin), 3.7 in the middle panel 
(SDSS observations per position, averaged over bin), and 5 (Trojans per 
deg$^2$, averaged over 4 deg$^2$ large bin). The purple (dark) regions 
contain no data. } 
\label{maps}
\end{figure*}

\begin{figure*} 
\centering
\includegraphics[width=17.5cm]{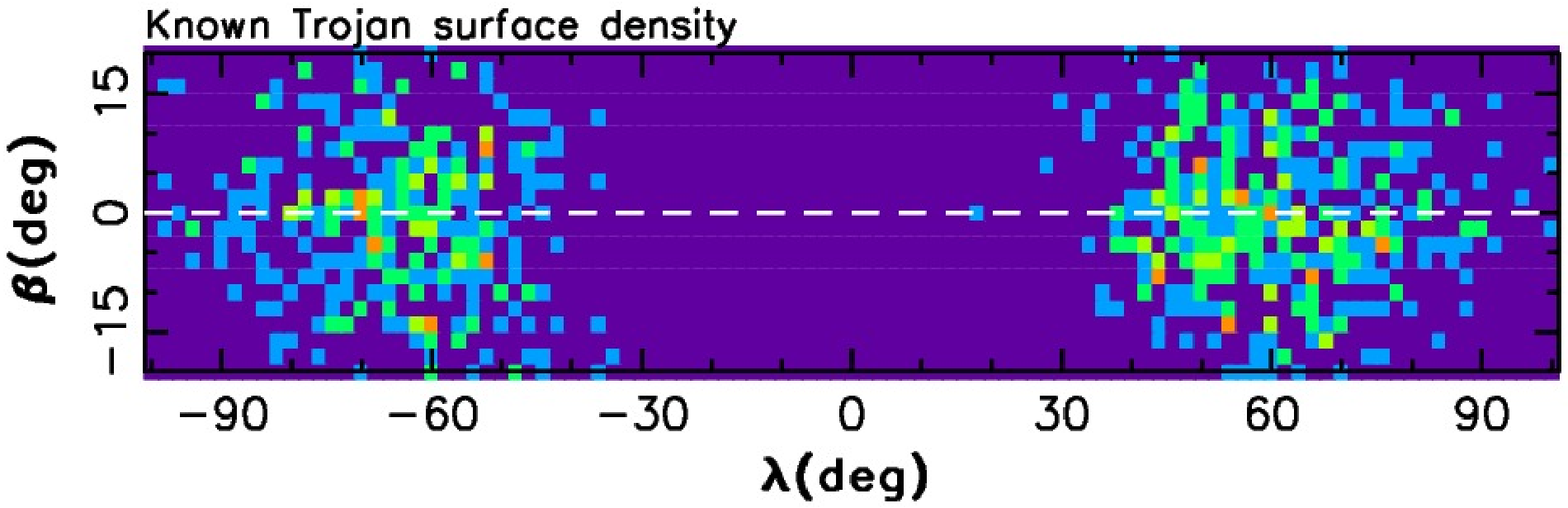}
\includegraphics[width=17.5cm]{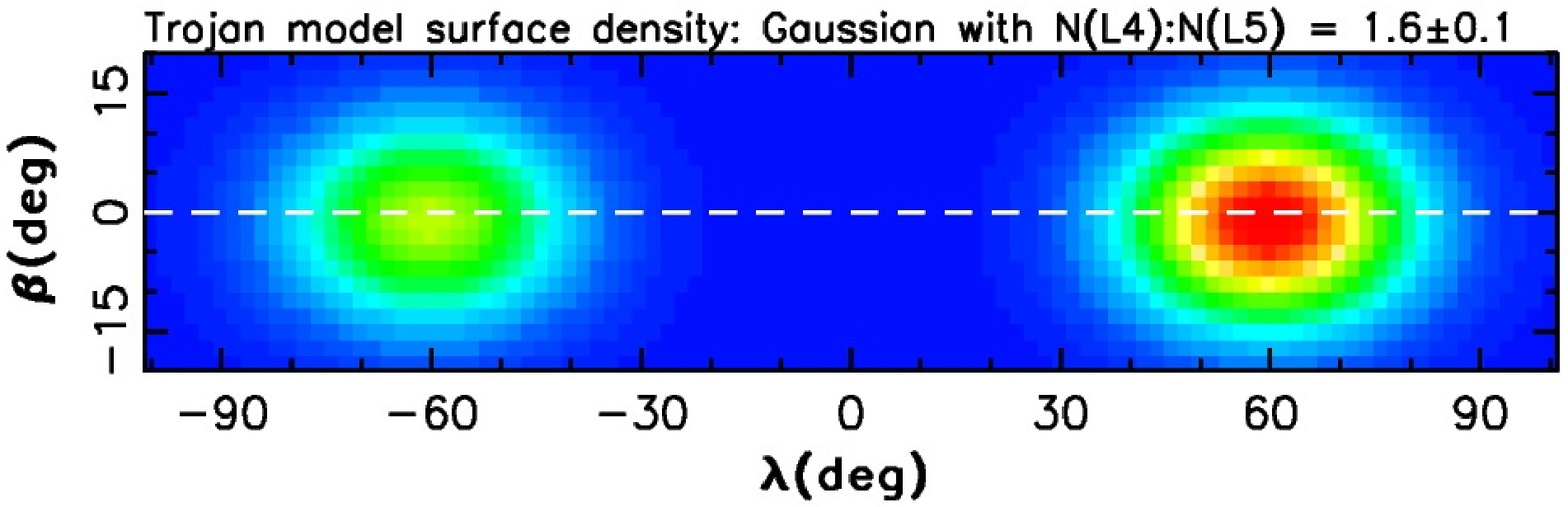}
\includegraphics[width=17.5cm]{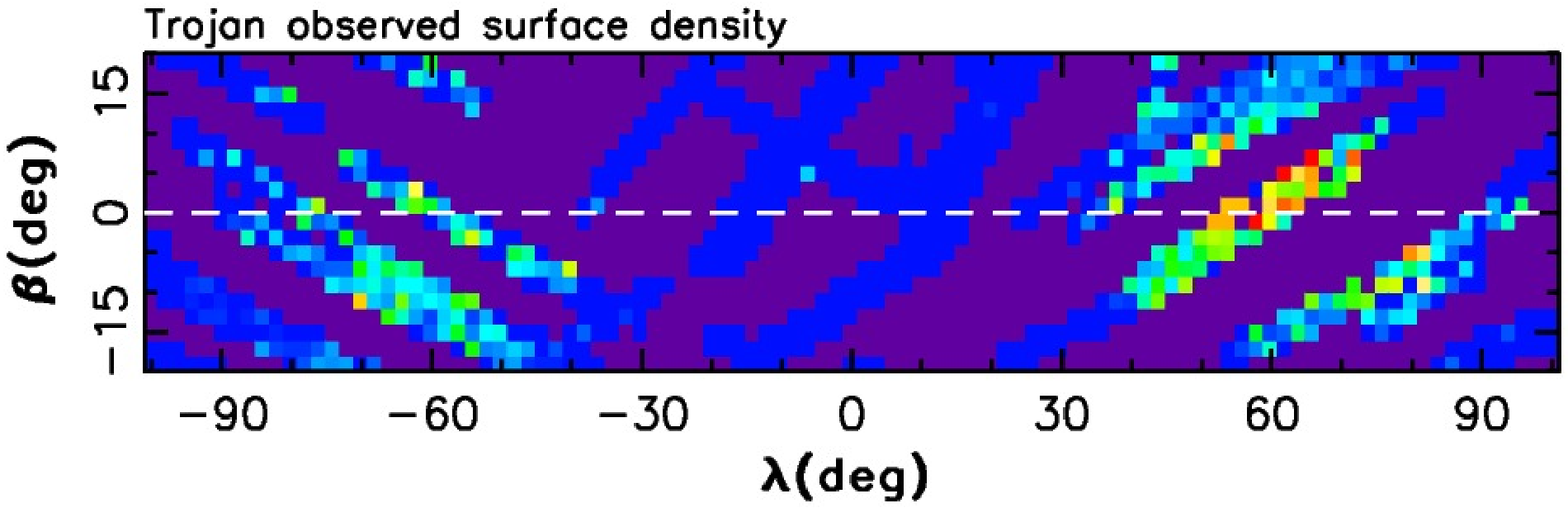}
\includegraphics[width=17.5cm]{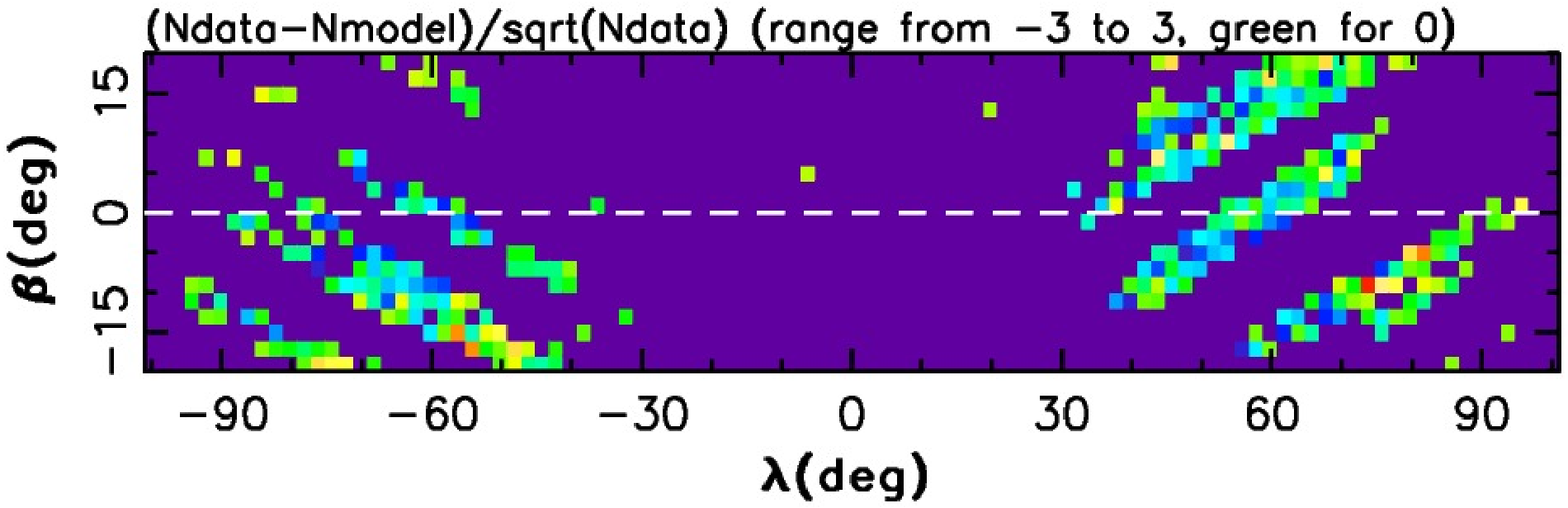}
\caption{
The top panel shows the observed surface density of known Trojan 
asteroids from Bowell's ASTORB file, analogously to Fig.~\ref{maps}.
The distribution for each swarm is well described by a two-dimensional 
Gaussian. The second panel shows a model distribution that has the  
same {\it shape} as the Gaussian distribution implied by the top panel, 
but normalized to the observed counts of SDSS candidate Trojan (for each 
swarm separately), shown in the third panel with the same color scheme
(red corresponds to 5 objects per deg$^2$). 
The best-fit L4:L5 number ratio is 1.6$\pm$0.1. The difference between 
the observed counts and this model distribution, normalized by the Poisson 
error bars, is shown in the bottom panel (the purple regions contain no 
data). The value of $\chi^2$ per degree of freedom is 1.15.}
\label{maps2}
\end{figure*}

\subsection{      The Numbers of Asteroids in L4 ad L5 Swarms   }
\label{NL4vsL5}

It is usually assumed that the leading (L4) and trailing (L5) swarms
contain similar number of asteroids down to the same size limit (e.g. JTL). 
Although the number of {\it known} objects in L4 and L5 differ (e.g. as 
listed in Bowell's ASTORB file), this asymmetry 
is usually dismissed as due to complex selection biases in the sample
of Trojans with known orbits (e.g. Marzari et al. 2001). On the other hand, 
P\'al \& S\"uli (priv. comm.) find using numerical simulations that the 
perturbations by Saturn produce different stability regions for L4 and L5. 
This effect is suspected to cause about a factor of 2 population size difference 
between the leading and trailing swarms. Therefore, it seems worthwhile
to examine the number ratio for the two swarms implied by the SDSS data.

The top panel in Figure~\ref{maps} shows the observed surface density 
map of candidate Trojan asteroids in Jupiter's coordinate system.
There are 1.9 times more objects with  $\lambda_{Jup} > 0$ than with
$\lambda_{Jup} < 0$ (this asymmetry is already easily discernible in 
histograms shown in Figure~\ref{lambdaTest}). However, this map does 
not reveal true density distribution because selection biases are strong 
even when using a homogeneous survey such as SDSS. The most important 
selection effect is the varying number of SDSS observations as a function of 
position relative to Jupiter, as shown in the middle panel\footnote{Here 
we assume that the depth of SDSS imaging is constant, which is true to 
within several tenths of a magnitude.}. It is the ratio of these two maps 
that is the best estimate of the underlying distribution of Trojan asteroids. 
This map is shown in the bottom panel in Figure~\ref{maps}. 

It is still not straightforward to use the counts from this corrected map 
to assess the number-count ratio for the two swarms. The reason is that
the SDSS coverage of the $\lambda_{Jup}-\beta_{Jup}$ plane is not 
symmetric with respect to  $\lambda_{Jup}=0$, and thus the counts cannot
be simply summed up and compared. At the same time, the shape of the
underlying distribution in the $\lambda_{Jup}-\beta_{Jup}$ plane is not known.

We use two different methods to solve this problem. The first one assumes that 
the {\it shape} of the true distribution of Trojan asteroids in the 
$\lambda_{Jup}-\beta_{Jup}$ plane is symmetric with respect to $\lambda_{Jup}=0$ 
and $\beta_{Jup}=0$, and the second one estimates this shape using a sample 
of known Trojans and normalizes it using the CT sample. 

Although the coverage of the $\lambda_{Jup} - \beta_{Jup}$ plane
by the available observations is fairly sparse, there is sufficient 
overlap of regions with the same $|\beta_{Jup}|$ and $|\lambda_{Jup}|$ 
to compute the number ratio for the two swarms. With the assumption 
of symmetry with respect to $\lambda_{Jup}=0$ and $\beta_{Jup}=0$, we
determine that the leading-to-trailing number ratio is 1.8$\pm$0.2 
(weighted average of all pixels). {\it It appears that the leading swarm 
has almost twice as many objects as the trailing swarm.}

The accuracy of the number-count ratio estimate can be increased when
the shape of the $\lambda_{Jup}-\beta_{Jup}$ distribution is assumed to 
be known (because all the data are used). We determined this shape using
a sample of 1178 known Trojan asteroids from ASTORB file. Their distribution in
the $\lambda_{Jup}-\beta_{Jup}$ plane is shown in the top panel in 
Figure~\ref{maps2}. We find that the shape of this distribution is well 
described by two two-dimensional Gaussians centered on $\beta_{Jup}=0$ deg
and $\lambda_{Jup}=\pm60$ deg, with the widths ($\sigma$) of 9 deg and 14 deg,
for $\beta$ and $\lambda$, respectively\footnote{The errors for these estimates
are not larger than $\sim0.5$ and indicate that the distribution of Trojans
on the sky is not circularly symmetric around L4 and L5 points, as assumed
by JTL.}. Using this shape, we fit the 
overall normalization for each swarm separately (i.e. two free parameters)
and obtain the leading-to-trailing number ratio of 1.6$\pm$0.1. The 
best-fit model is shown in Figure~\ref{maps2}, as well as the residual
map. With the
assumption that the $\lambda_{Jup}-\beta_{Jup}$ distribution does not
depend on size, this is our best estimate for the relative number count
normalization for the two swarms. It is reassuring that we obtained a 
statistically consistent result using the first method. We emphasize
that there is no discernible difference in the shape of the spatial
distribution of objects from the two swarms. 

Interestingly, this number ratio is about the same as the 
leading-to-trailing ratio of Trojans with known orbits in Bowell's 
ASTORB file. Although the selection effects are typically invoked to 
explain this asymmetry, it instead appears to be a real effect (we show 
below that the sample of known Trojans is indeed fairly complete to 
$r\sim19.5$). On the other hand, the number ratio of asteroids in the two 
swarms could be dependent on object's size, and the SDSS sample extends
to smaller sizes than ASTORB file. We address this possibility 
in the next section.

\subsection {     Apparent and Absolute Magnitude Distributions      }
\label{magDist}

\begin{figure} 
\centering
\includegraphics[bb=140 200 515 761, width=8cm]{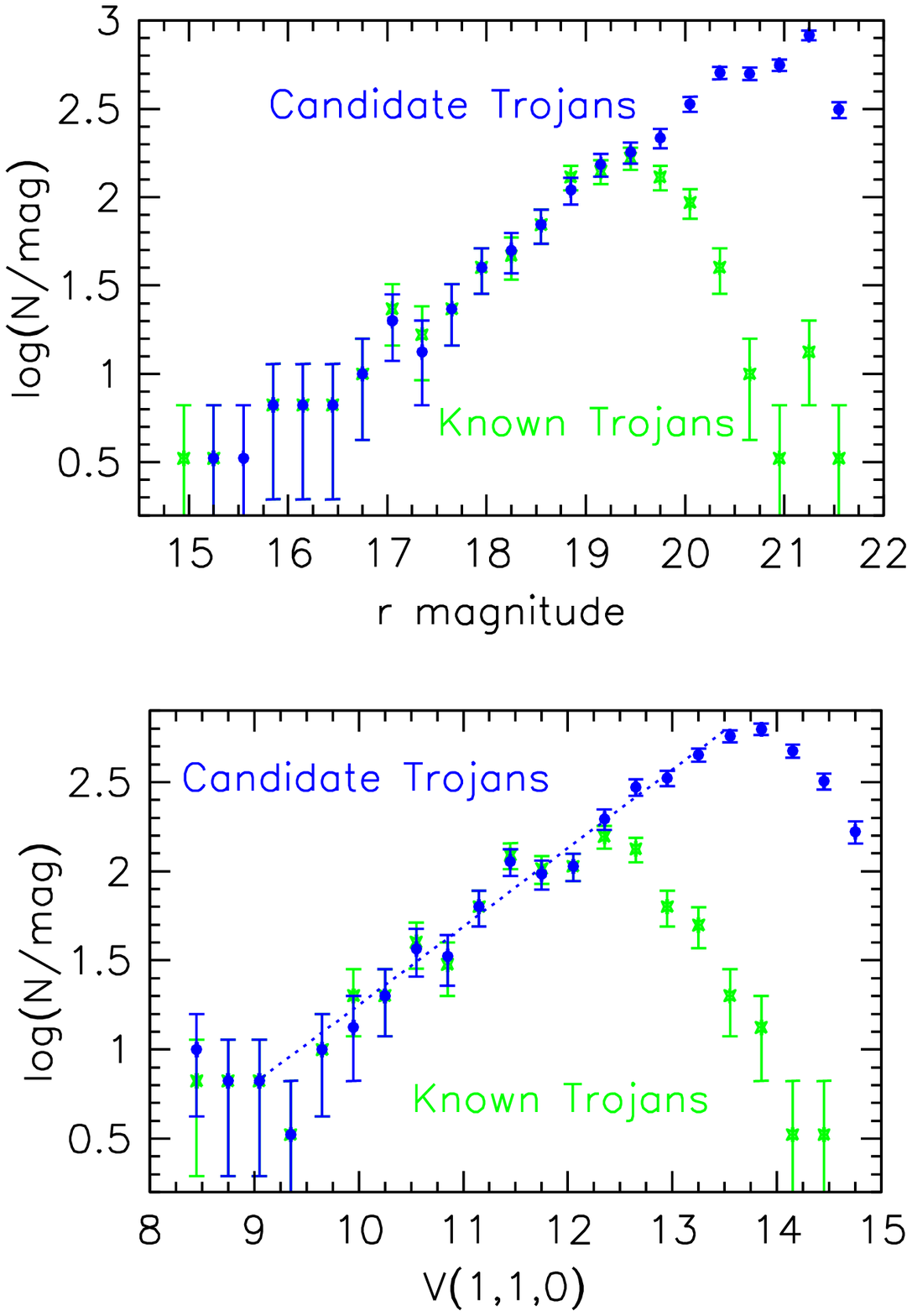}
\caption{The top panel shows the differential SDSS $r$ band distributions 
for known (squares) and candidate Jovian Trojan asteroids (circles). 
The SDSS candidate sample is $\sim$1.5 mag deeper than the sample of known 
objects. The bottom panel compares the differential absolute magnitude
distributions in the Johnson's $V$ band. The dashed line is added to guide
the eye and has the slope of 0.44. The SDSS data suggest that practically 
all Trojans brighter than $V(1,1,0)\sim12.3$ ($r\sim19.5$), or approximately larger 
than 20 km, are already discovered and listed in ASTORB file.}
\label{magDistFig}
\end{figure}

\begin{figure} 
\centering
\includegraphics[bb=134 344 510 670, width=8cm]{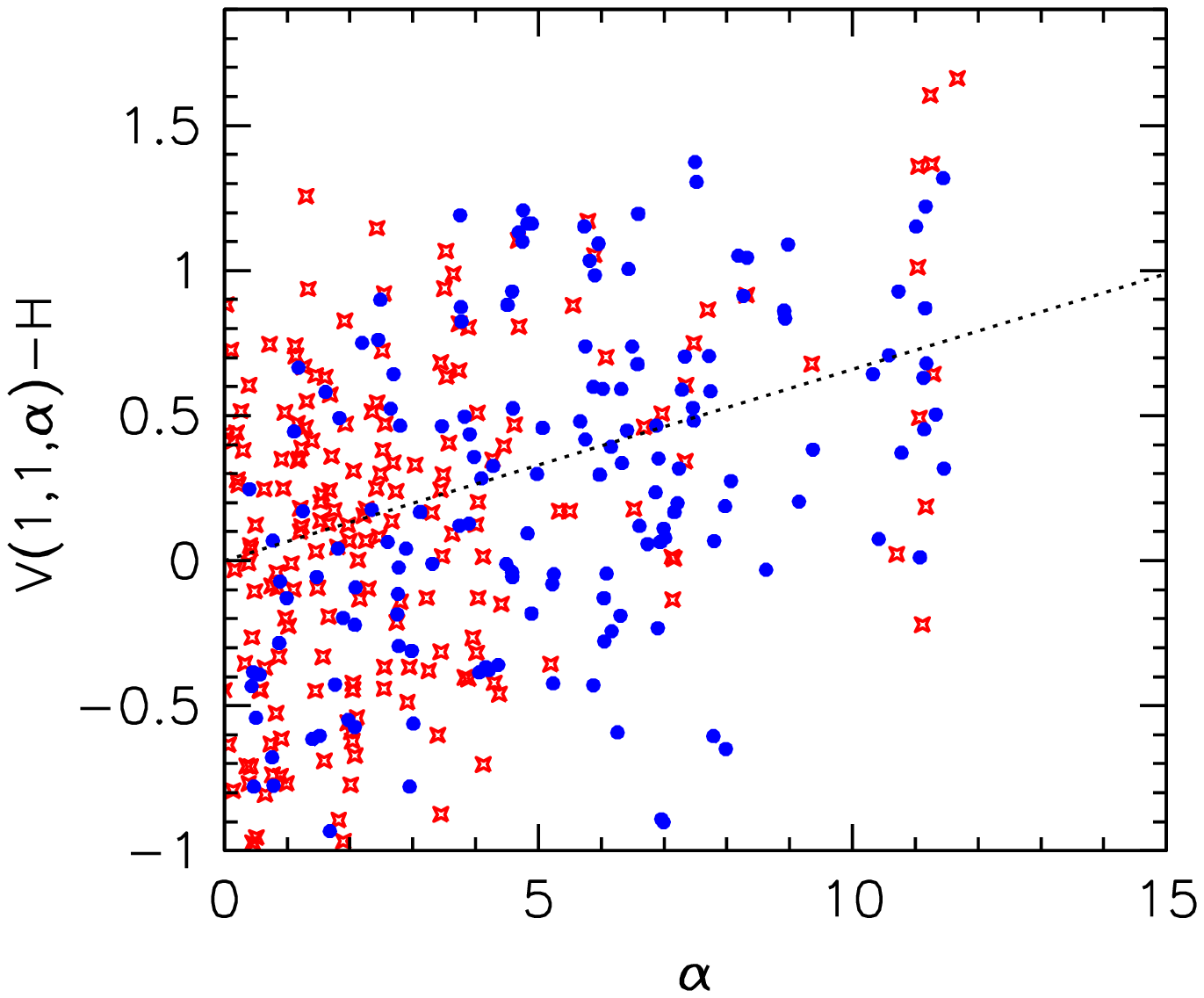}
\includegraphics[bb=184 224 560 570, width=8cm]{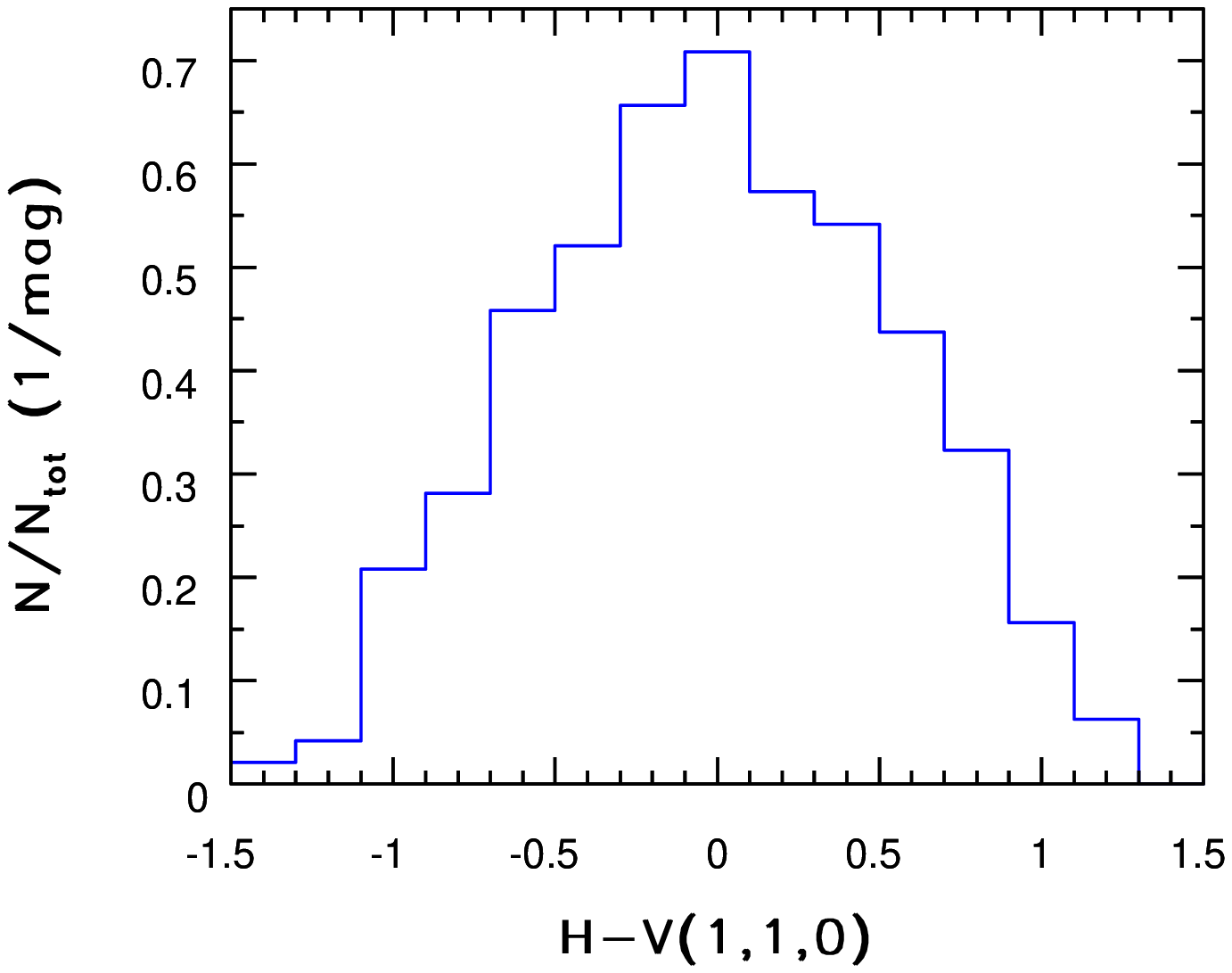}
\caption{The calibration of phase effects on observed magnitudes. The top
panel shows the distance-corrected magnitudes as a function of phase for known
Trojans observed at small latitudes ($|\beta |<10$). Two different symbols corresponds
to objects from L4 (star) and L5 (dot) swarms. The dotted line shows
a best linear fit discussed in the text. The bottom panel shows a histogram 
of the scatter around this best fit.
}
\label{PhaseCoeff}
\end{figure}

\begin{figure} 
\centering
\includegraphics[bb=140 110 515 545, width=8cm]{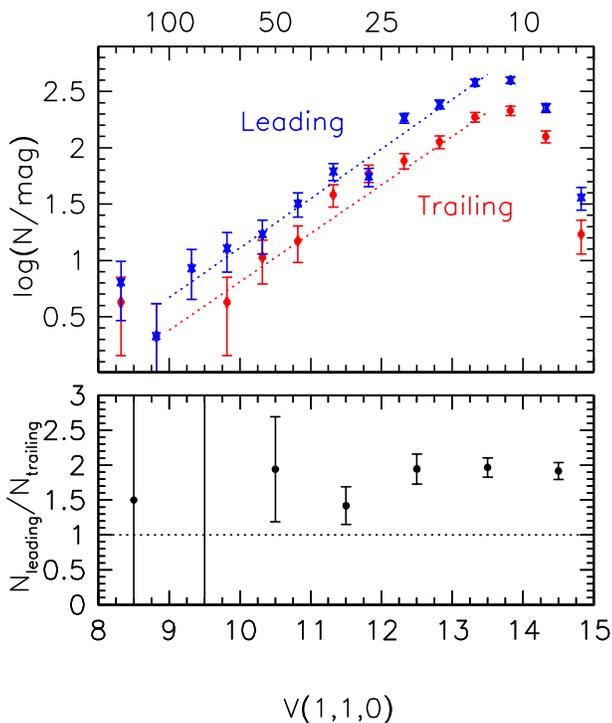}
\caption{The top panel compares the {\it differential} distributions of 
estimated absolute 
magnitudes in the Johnson $V$ band for candidate Trojans separated into
leading and trailing swarms. The counts have indistinguishable slopes,
but the overall normalization is different. The bottom panel illustrates
this difference by showing the ratio of {\it cummulative} counts for the
two swarms. Note that within errors this ratio does not depend on absolute
magnitude, or equivalently size, as marked on top (diameter in km). }
\label{L4L5counts}
\end{figure}

\begin{figure} 
\centering
\includegraphics[bb=130 110 505 545, width=8cm]{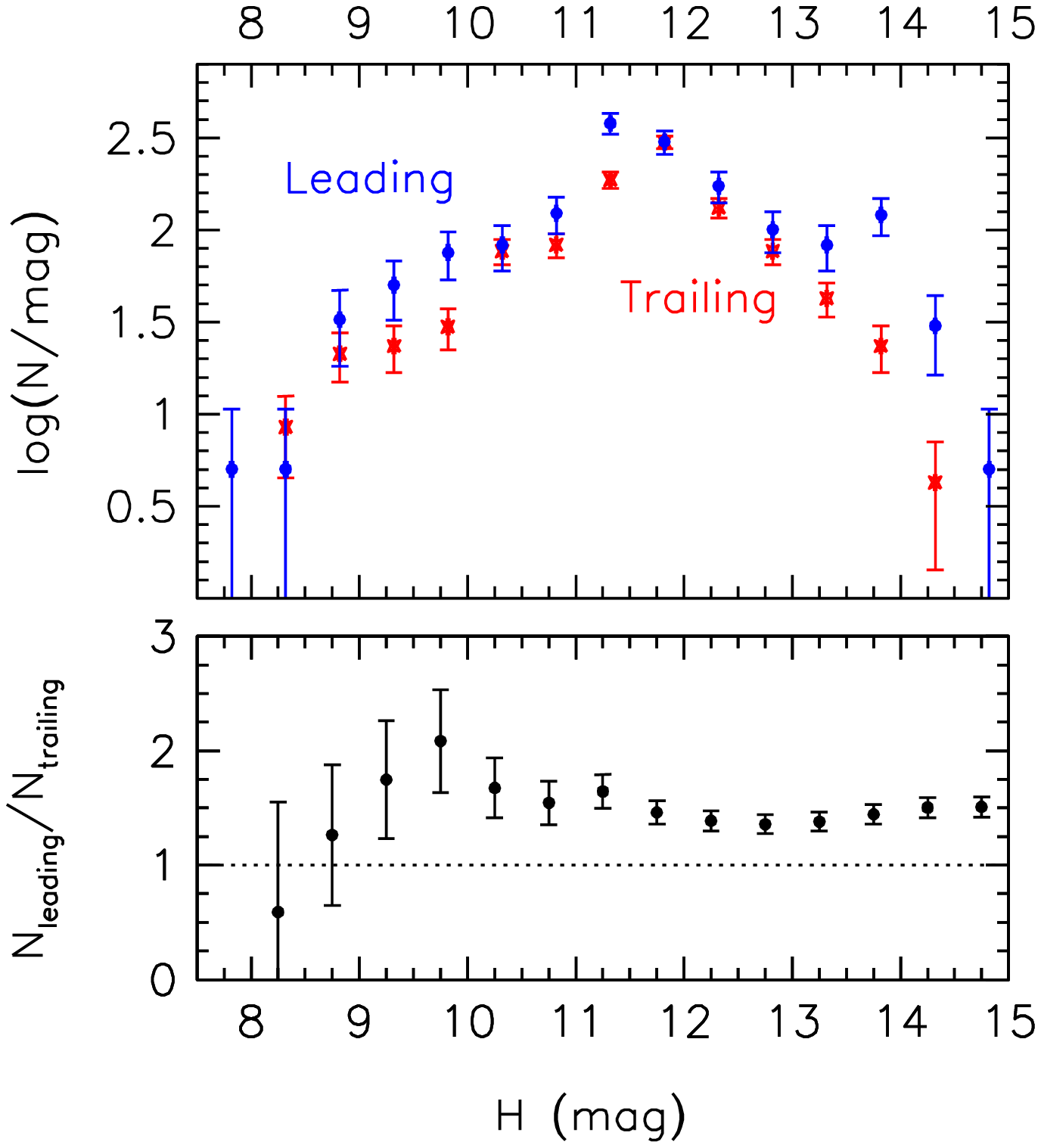}
\caption{Analogous to Fig.~\ref{L4L5counts}, except that {\it all} known 
Trojans listed in Bowell's ASTORB file are included, and that absolute magnitude 
estimator $V(1,1,0)$ is replaced by the measured value $H$. Note that 
the leading swarm has $\sim$1.5 times more objects than the trailing swarm
at the completeness limit ($H\sim12$).
}
\label{L4L5countsAllBowell}
\end{figure}

The differential apparent $r$ band magnitude distributions (for Trojans,
Johnson's $V \sim r + 0.25$) for known (KT) and candidate (CT) Trojans are 
shown in the top panel in Figure~\ref{magDistFig}. The KT sample is complete to 
$r\sim19.5$, and the CT sample is complete to $r\sim21$. The formal
cutoff for inclusion of moving objects in the SDSS MOC is $r<21.5$. 
A slightly brighter completeness limit for Trojans can be understood
as the removal of objects from a fairly narrow velocity space due 
to velocity errors (see fig.~6 from I01). Because the CT sample is complete 
to a $\sim1.5$ mag deeper limit, it contains $\sim$4 times more objects. 
It is noteworthy that the high completeness
of KT sample indicated by the SDSS data (that is, the counts are 
practically identical for $r<19.5$) argues that selection effects 
{\it cannot} be invoked to explain the L4--L5 asymmetry in the number
counts of Trojans with known orbits listed in Bowell's ASTORB file. 

In order to investigate the dependence of various quantities (such as
counts and colors) on object size, we transform apparent magnitudes
to absolute magnitudes as follows. The dependence of apparent magnitude
in the Johnson $V$ band on absolute magnitude, $H$, distance from Sun,
$R$, distance from Earth, $\Delta$, and viewing (phase) angle, $\alpha$, 
can be expressed as 
\begin{equation} 
    V(R,\Delta,\alpha) = H + 5\log(R\Delta) + F(\alpha).
\end{equation} 
Here $V(R,\Delta,\alpha)=r + 0.44(g-r)$ is synthesized from SDSS
measurements, $F(\alpha)$ is the phase function, and $H$ includes the 
dependence on diameter $D$ (in km) and the V-band albedo, $p_V$
\begin{equation} 
         H = 19.14 - 2.5\log({p_V \over 0.04}) - 5\log(D).
\end{equation} 
Note that formally $V(1,1,0)=H$. Given $R$ and $\phi$, $\Delta$ and $\alpha$
can be found from 
\begin{equation} 
    \Delta^2 + 2 \Delta \cos(\phi) + 1 = R^2
\end{equation} 
and
\begin{equation} 
    \alpha = \phi - \arccos \left( {1 + \Delta \cos(\phi) \over R} \right).
\end{equation} 

When applying this procedure to observations discussed here, $R$ and 
$F(\alpha)$ are not known. We adopt\footnote{I01 developed a method for 
estimating heliocentric distance of asteroids from their angular velocity
measured by SDSS that is accurate to about 10\% for main-belt asteroids.
For Trojans, which have larger velocity errors, a smaller error is introduced
by assuming a constant $R$.} $R=5.2$ AU and model the phase function
as $F(\alpha)= k |\alpha|$. Therefore,
\begin{equation} 
      H \sim V(1,1,0) = V(1,1,\alpha) - k |\alpha|.
\end{equation} 

In order to determine coefficient $k$, we used known Trojan asteroids observed 
at low latitudes ($|\beta|<10$). 
A least-square best-fit to the observed dependence of $V(1,1,\alpha)-H$ on $|\alpha|$
gives $k=0.066\pm0.018$. To the zero-th order, the transformation from 
apparent to absolute magnitudes for Trojans observed close to the opposition
amounts to a shift of about 7 mag. In order to distinguish absolute magnitude for 
objects with known orbits from the estimates evaluated here, we will refer 
to $H$ and $V(1,1,0)$ for KT and CT samples, respectively. 

It is noteworthy that the intercept of the best-fit line discussed above
(see  Figure~\ref{PhaseCoeff})
is consistent with 0. This shows that the $V$ band magnitudes synthesized from
SDSS photometry and $H$ magnitudes for Trojans listed in ASTORB file are 
expressed on the same photometric system. This appears not be the case for 
a significant fraction of main-belt asteroids whose magnitudes (that are
simply adopted from a variety of asteroid surveys) can have systematic 
errors as large as 0.5 mag (for more details see J02). The root-mean-square
width of the residuals distribution shown in the bottom panel in Figure~\ref{PhaseCoeff}
is 0.3 mag, and represents an upper limit for the errors of our method for
estimating $V(1,1,0)$ (e.g. photometric and other errors for $H$ listed in 
ASTORB file and intrinsic
object variability probably also contribute).

The bottom panel in Figure~\ref{magDistFig} compares the differential absolute 
magnitude distributions in the Johnson's $V$ band for known and candidate Trojans. 
The SDSS data suggest that practically all Trojans brighter than
$V(1,1,0)\sim12.3$, or those with diameters approximately larger than 20 km, 
are already discovered.

We find that the differential absolute magnitude distribution is well described by
\begin{equation} 
            \log(N) = C +  \alpha H
\end{equation} 
with $\alpha=0.44\pm0.05$. This implies a differential size distribution index 
of $q=5\alpha + 1 = 3.2 \pm 0.25$, valid for $9<H<13.5$. 
This value is in good agreement with JLT, who obtained 
$q=3.0\pm0.3$ using about 10 times smaller sample, and with Yoshida \&
Nakamura (2005), who obtained $q=2.9\pm0.1$ using a sample of 51 objects. 

We use the counts of the presumably complete bright ($H<12$, see 
Figure~\ref{magDistFig}) subsample of
known Trojans from ASTORB file to normalize the {\it cumulative} counts
\begin{equation} 
\label{Ncum}
            \log(N_{cum}) = 2.9 + 0.44\,(H-12)
\end{equation}  
Assuming $p_V$=0.04 (Fern\'{a}ndez, Sheppard \& Jewitt 2003), $D$=1 km 
corresponds to $H$=19.14. The above result
implies that there are about 1 million Jovian Trojans larger
than 1 km, to within a factor of 2 (uncertainty comes from the
error in $\alpha$ and extrapolation over 5~mags; in addition, this 
normalization scales with the albedo approximately as $\propto 0.04/p_V$). 
This estimate could be up to a factor of 2 too high if Trojans size 
distribution becomes shallower for objects smaller than $\sim$5~km, as 
was found for main-belt asteroids (see I01), and is suggested for
Jovian Trojans by Yoshida \& Nakamura (2005). These results are in 
good agreement with the normalization obtained by JTL and imply that
{\it there are about as many Jovians Trojans as there are main-belt 
asteroids down to the same size limit.}

In previous section, we demonstrated that L4 has a significantly larger
number of objects than L5. To examine whether, in addition to this
difference in overall normalization, the {\it slope } of the size 
(i.e. $H$) distribution is different for the two swarms, we separately 
analyzed their counts. The slope of the size distributions for both the 
candidate Trojans (Figure~\ref{L4L5counts}) and for known Trojans 
(Figure~\ref{L4L5countsAllBowell}) are the same within measurement 
uncertainties (with the slope error $\sim$0.05). Note that the
L4-to-L5 count ratios shown in the bottom panel in Figures~\ref{L4L5counts}
and Figure~\ref{L4L5countsAllBowell} are different from the value of 
1.6 discussed in Section~\ref{NL4vsL5} because $\lambda - \beta$ selection 
effects (which are not a function of size) are not taken into account.

\subsection{              Color Distribution                 }

\begin{figure*} 
\centering
\includegraphics[bb=19 406 570 779, width=15cm]{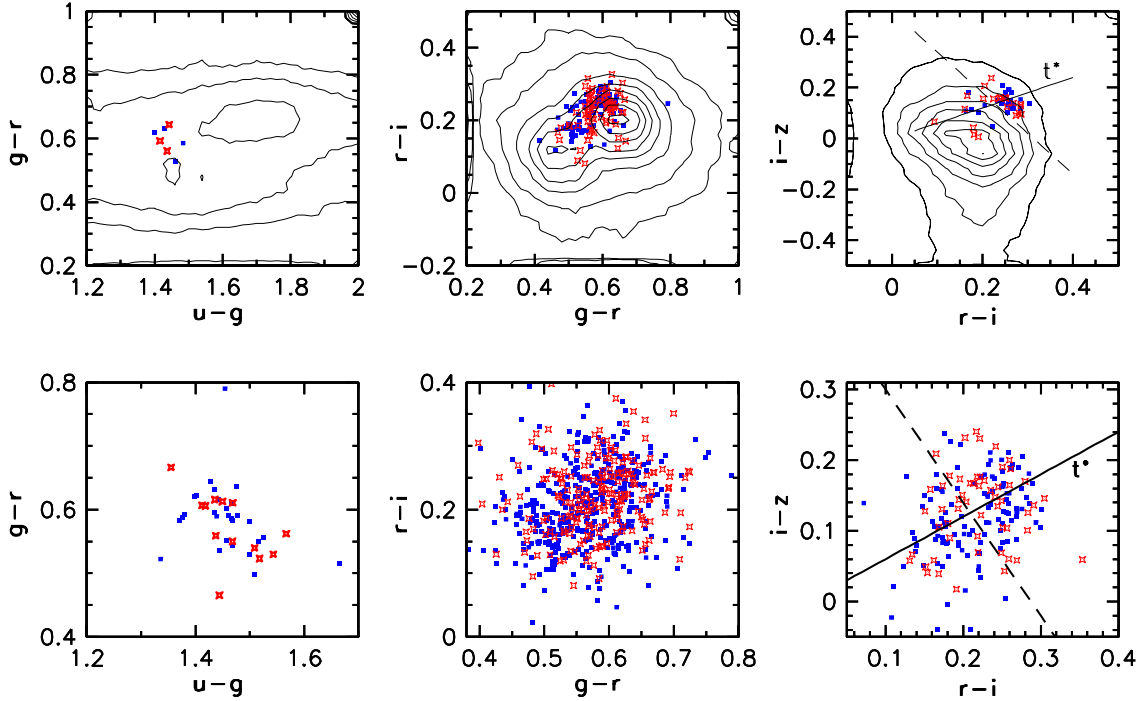}
\caption{The top panels compare the color distribution of the known Trojan 
asteroids (symbols, blue for L4 and red for L5 swarm) to the color 
distribution of main-belt asteroids (contours, linearly spaced). The bottom 
panels zoom in on the distribution of candidate Trojans, which is similar
to that of the known Trojans. Measurement errors are typically less than  
0.05 mag. The two lines in the $i-z$ vs. $r-i$ diagrams shown on the right
define principal colors (see text).}
\label{CCDfig}
\end{figure*}

\begin{figure} 
\label{albedo}
\centering
\includegraphics[bb=140 190 465 605, width=8cm]{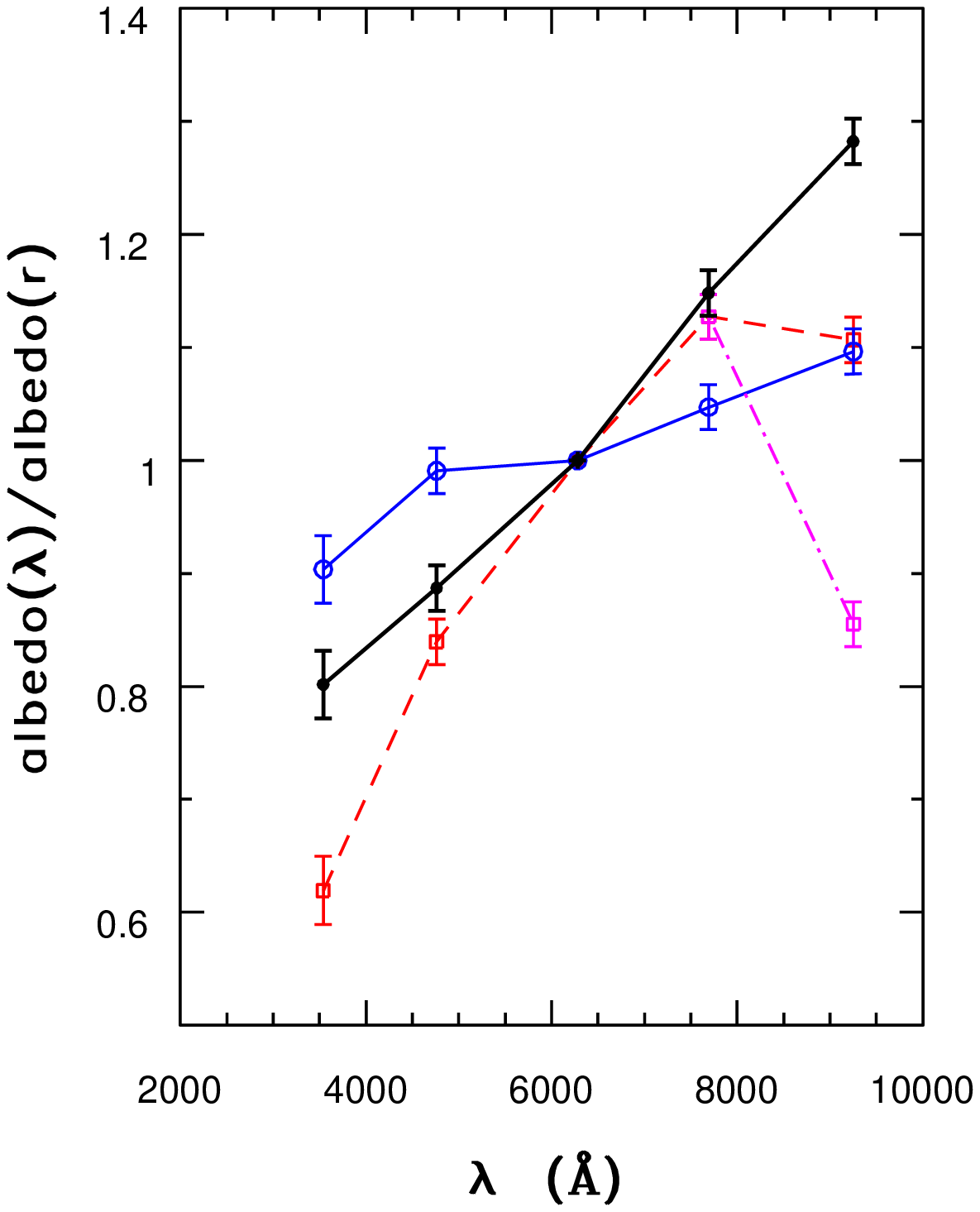}
\caption{A comparison of the relative albedo for Trojan asteroids
(black dots) and the relative albedo for the three dominant main-belt 
color types (C type: blue circles, S type: red solid squares, V type:
magenta open squares, for bands other than $z$ same as S). Due to large 
sample sizes, errors reflect systematic uncertainties in SDSS photometric 
calibration.}
\label{albedo}
\end{figure}

One of the main advantages of the sample discussed here are
accurate color measurements for a sample about two orders of 
magnitude larger than available before. Together with robust
knowledge about the color distribution of main-belt asteroids
in the SDSS photometric system (I01, I02a), we are in a position
to compare the colors of the two populations with an unprecedented
level of detail. 

We first correct color measurements for the phase effects using
a linear color vs. phase angle approximation discussed in 
Section~\ref{magDist}. We obtained the following best-fit relations 
for the colors corrected to zero phase angle
\begin{equation} 
    (g-r)_c  = (g-r) - 0.0051 \, |\alpha|,
\end{equation} 
and
\begin{equation} 
    (r-i)_c  = (r-i) - 0.0056 \, |\alpha|,
\end{equation} 
with the coefficient errors of about 0.001 mag/deg. No significant 
correlation with the phase angle was detected for the $i-z$ color,
and too few objects have accurate $u-g$ color measurement to attempt
a robust fit. As the mean value of $|\alpha|$ is about 2 degree, these
corrections are small compared to photometric accuracy. 

The color distribution of Trojan asteroids is compared to the color
distribution of main-belt asteroids in Figure~\ref{CCDfig}. The mean 
colors and their standard deviation (not the error of the mean!) 
for candidate Trojans with color errors less than 0.05 mag are 
$u-g=1.45, 0.08$, $g-r=0.55, 0.08$, $r-i=0.22, 0.10$, and $i-z=0.13, 
0.11$ (for reference, these colors correspond to Johnson's $B-V=0.73$,
$V-R=0.45$, and $R-I=0.43$, using the photometric transformations 
from Ivezi\'{c} et al. 2007; these values are in good agreement with
previous work, e.g., Fornasier et al. 2004, Dotto et al. 2006).
The two distributions are different, with 
the difference maximized in the $i-z$ vs. $r-i$ diagram. Using solar colors 
from I01, we compute the relative albedo for Trojan asteroids and compare
it to the three dominant main-belt color types in Figure~\ref{albedo}.
As expected from previous work, Trojan asteroids are redder than 
main-belt asteroids at wavelengths longer than the visual band.

In addition to maximizing color differences between Trojan and main-belt 
asteroids, the $i-z$ vs. $r-i$ diagram is interesting because the
distribution of candidate Trojans suggests bimodality. To quantify 
this effect in the subsequent analysis, we define a color index which 
is a linear combination of the $r-i$ and $i-z$ colors:
\begin{equation}
   t^* = 0.93\,(r-i) + 0.34\,(i-z) -0.25,
\end{equation}
with the phase-angle correction 
\begin{equation}
  t_c^* = t^* - 0.005 \, |\alpha|.
\end{equation}

The distribution of this color index for known and candidate Trojans is 
compared to that of the main-belt asteroids in Figure~\ref{tColorHist}. 
The fact that the distributions for known and candidate Trojans are 
indistinguishable, while clearly different from that of the main-belt 
asteroids, is another demonstration of the robustness of kinematic 
selection method.  

The $t^*$ distribution for Trojan asteroids is bimodal. At first it 
appears that this bimodality is related to L4 vs. L5 separation, as 
illustrated in Figure~\ref{tColorHist}. However, objects from L4 and L5 
have different {\it observed} orbital inclination distribution due to 
observational selection effects (see Section~\ref{colorInc}). Instead, 
the differences in the L4 and L5 color distributions are due to 
a color-inclination correlation, as detailed below.

\subsubsection{ Correlation between Color and Orbital Inclination }
\label{colorInc}

\begin{table*}
\begin{tabular}{rlllllllllllllrr}
$inc$ & N & 
$\langle$g$-$r$\rangle$&Err& 
$\langle$r$-$i$\rangle$&Err& 
$\langle$i$-$z$\rangle$&Err& 
$\langle$B$-$V$\rangle$&Err& 
$\langle$V$-$R$\rangle$&Err& 
$\langle$R$-$I$\rangle$&Err&
$\langle$t$\rangle$&Err \\
\hline
0--10 & 153 & 
0.56 & 0.01 & 0.21 & 0.01 & 0.11 & 0.02 &  0.73 & 0.02 & 0.45 & 0.01 & 0.42 & 0.01 & $-$0.02 & 0.01\\
10--20 & 227 & 
0.58 & 0.01 & 0.24 & 0.01 & 0.13 & 0.01 &  0.75 & 0.01 & 0.47 & 0.01 & 0.47 & 0.01 & 0.01 & 0.01\\
20--30 & 71 & 
0.60 & 0.02 & 0.26 & 0.01 & 0.16 & 0.01 &  0.77 & 0.02 & 0.48 & 0.01 & 0.48 & 0.01 & 0.04 & 0.01\\
\hline
\end{tabular}
\caption{The statistics of various color indices show prominent inclination dependence. 
The subsets are selected by the inclination range, $inc$, $N$ is the number of objects in each bin,
and $Err$ is the standard error of the mean.}
\end{table*}

As was already discernible in Figure~\ref{pP1}, the color and 
orbital inclination for Jovian Trojan asteroids are correlated. This
correlation is presented in a more quantitative way in the top panel
in Figure~\ref{FigIncColor} and in Table 1. As evident, objects with large orbital
inclination tend to be redder. For example, the median $t^*$ color is 
-0.01 for objects with inclination less than 10 degree, while it is
0.04 for objects with inclination greater than 10 degree, and 0.06
for those with inclination greater than 20 degree. While these 
differences are not large, they are detected at a statistically 
significant level (the formal uncertainties are smaller than 0.01 mag). 
Equivalently, the median inclination for objects with $t^* < 0$ is
8.9 degree, while it is 13.4 for the redder objects. The marginal
color distributions for subsamples selected by inclination are
shown in the left panel in Figure~\ref{FigIncColorHist}. 

The sample of candidate Trojans is much larger and fainter than the 
sample of known Trojans and can be used to test whether the 
color-inclination correlation extends to smaller sizes. Since 
the orbital inclination is unknown for the majority of candidate
Trojans\footnote{I01 describe a method to estimate distance and
inclination of moving objects from their observed apparent motions.
While their method had satisfactory accuracy for studying main-belt
asteroids, we found using a simple Monte Carlo simulation that it 
is not applicable here because the three times slower apparent motion 
of Trojans results in unacceptably large inclination errors.}, 
we use as its proxy the latitude relative to Jupiter's 
orbit, $\beta$. When the sample of known Trojans is separated
by $\beta=6$ deg, 89\% of high-inclination and 66\% of low-inclination
objects are correctly classified. As evident from the middle panel 
in Figure~\ref{FigIncColorHist}, the differences in color histograms 
for subsamples of known Trojans separated by $\beta$ are still 
discernible, which justifies the use of $\beta$ as a proxy for inclination. 
The color histograms for candidate Trojans separated by $\beta$ are 
shown in the right panel in Figure~\ref{FigIncColorHist}. As they look 
similar to the analogous histograms for known Trojans, we conclude that 
the  color-inclination correlation extends to smaller sizes. 

\begin{figure*} 
\centering
\includegraphics[bb=36 438 570 645, width=15.5cm]{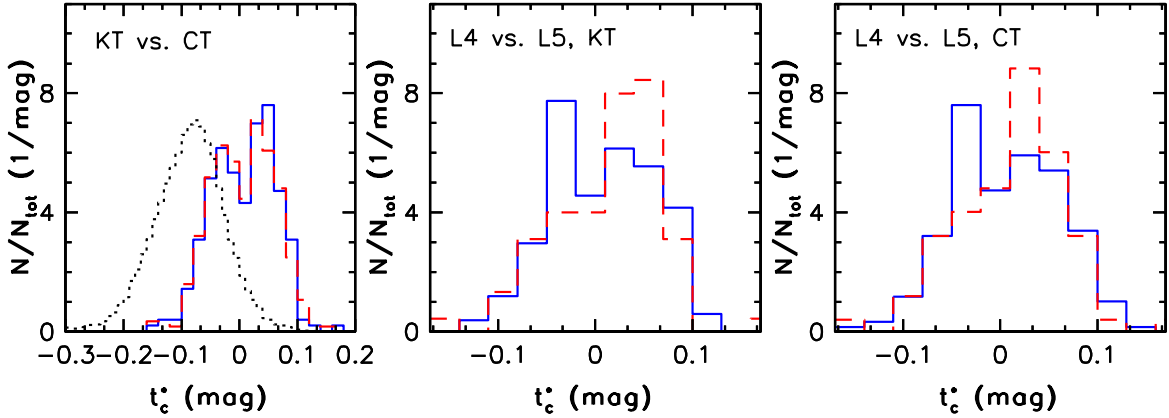}
\caption{The left panel compares the distribution of the synthetic
color index $t^*$ for known (dashed line) and candidate (solid line)
Trojan asteroids to that of the main-belt asteroids (dotted line). 
The middle and right panels compares the  $t^*$ distribution separately
for L4 (solid line) and L5 (dashed line) swarms. The differences 
between the two swarms are due to a color-inclination correlation
and different sampling of orbital inclinations due to observational
selection effects (see Section~\ref{colorInc}).}
\label{tColorHist}
\end{figure*}

\begin{figure} 
\centering
\includegraphics[bb=270 85 540 535, width=8cm]{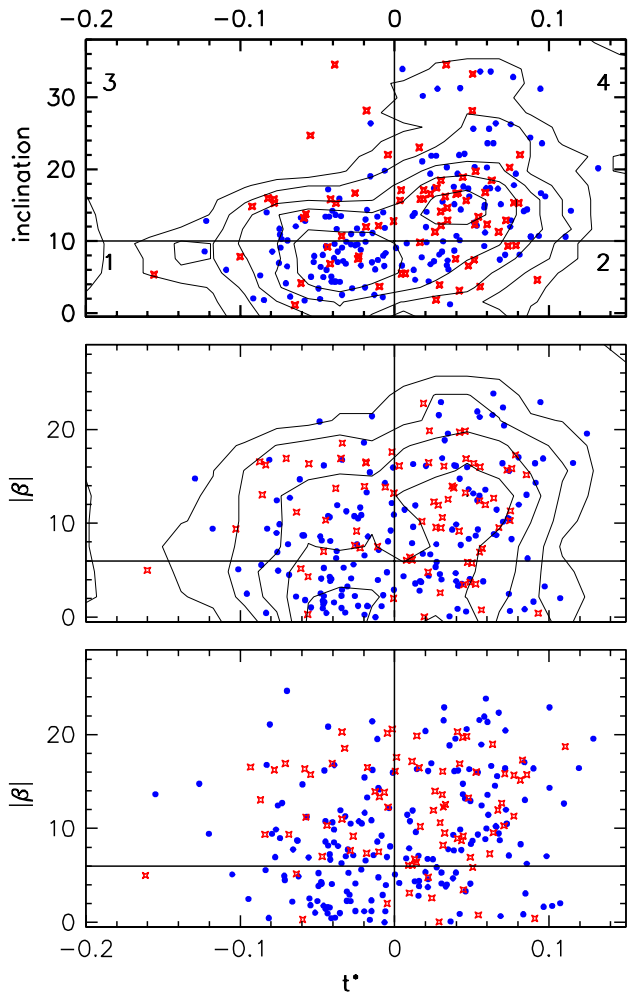}
\caption{The top panel shows the distribution of known Trojans in
the inclination vs. color diagram using linearly spaced contours.
Individual objects are also shown and separated into L4 (blue
dots) and L5 (red crosses) swarms. Note that L5 objects tend to
have larger inclination due to observational selection effects. 
The middle panel is analogous, except that orbital inclination 
is replaced by its proxy $\beta$ (latitude relative to Jupiter's 
orbit). The bottom panel is analogous to the middle panel, except 
that it shows candidate Trojans.}
\label{FigIncColor}
\end{figure}

\begin{figure*} 
\centering
\includegraphics[bb=36 438 570 645, width=15.5cm]{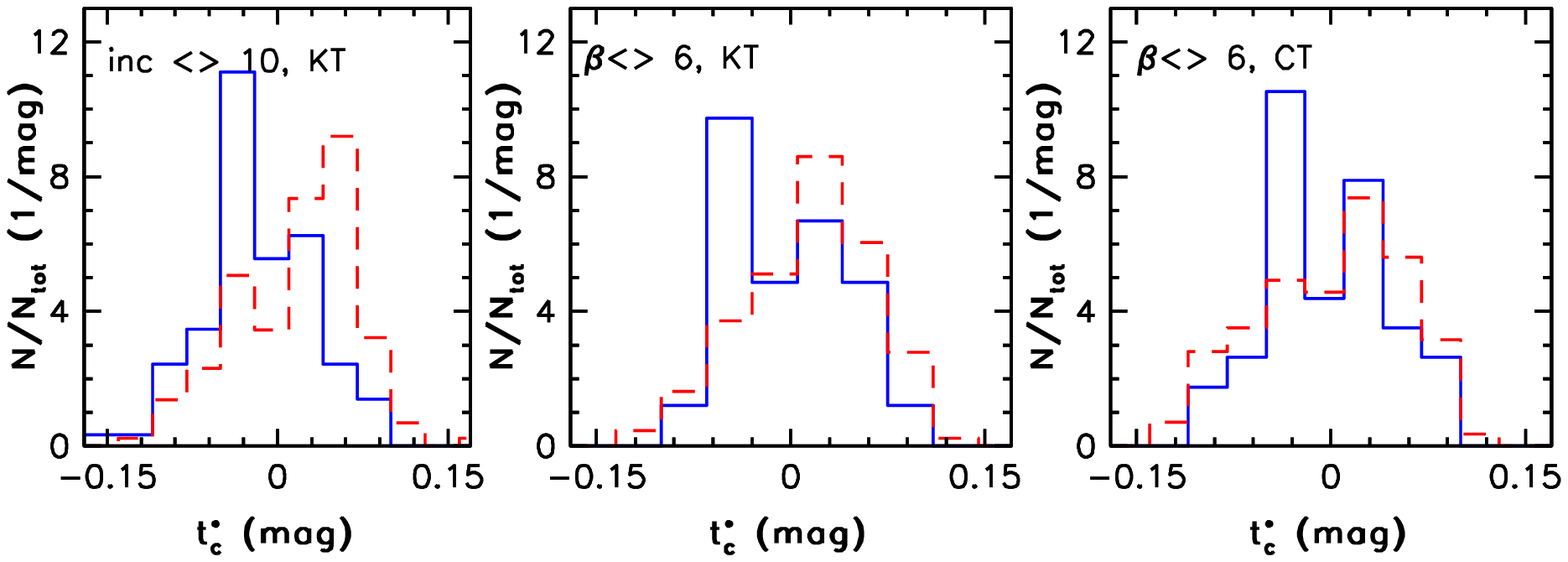}
\caption{The left panel compares the  $t^*$ color distributions for 
subsamples of known Trojans separated by orbital inclination ($<10$ deg:
solid line, $>10$ deg: dashed line. The middle panel is analogous,
except that subsamples are separated by the observed latitude relative 
to Jupiter's orbit ($\beta$). The right panel is analogous to the 
middle panel, except that it shows color distributions for candidate
Trojans. Note that objects with large inclinations and large $\beta$
tend to have redder colors.} 
\label{FigIncColorHist}
\end{figure*}

\begin{figure*} 
\centering
\includegraphics[bb=36 438 570 645, width=15.5cm]{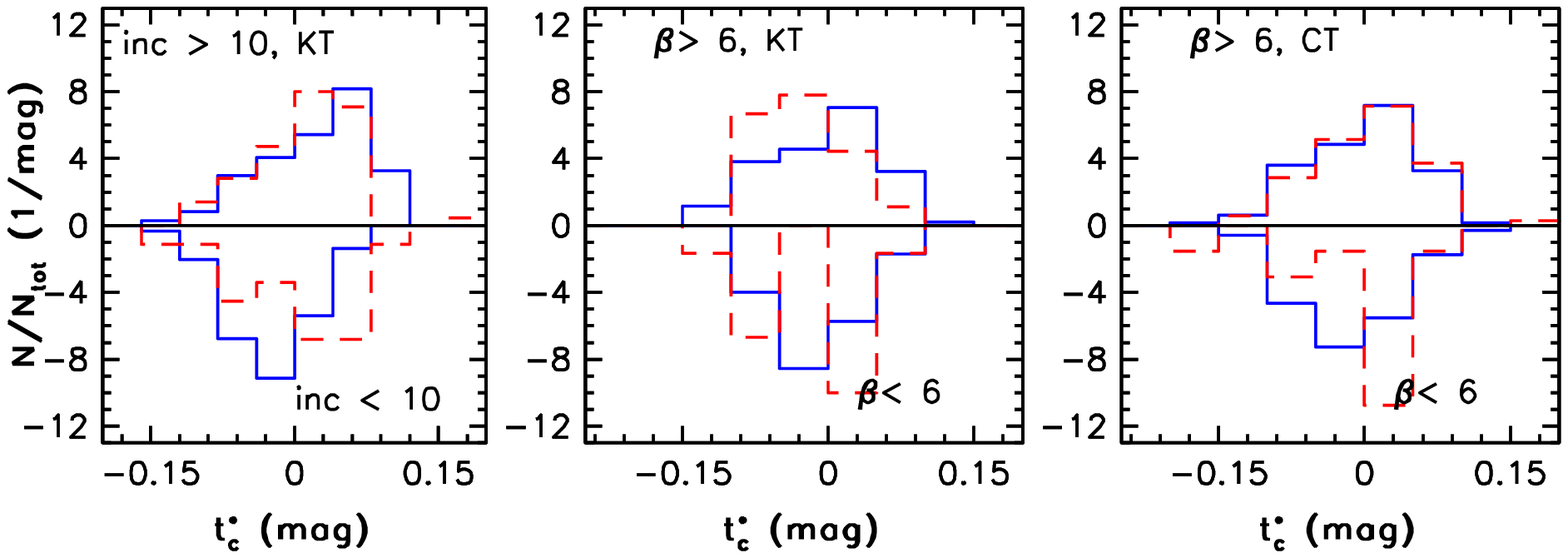}
\caption{Analogous to Figure~\ref{FigIncColorHist}, except
that each histogram is separated into contributions from 
each swarm. The top histograms correspond to histograms shown
by dashed lines in Figure~\ref{FigIncColorHist}, and the bottom
histograms to those shown by solid lines. Here solid line 
histograms correspond to L4 swarm and dashed line histograms
to L5 swarm. As evident, once the objects are separated by
inclination, or by $\beta$, the color difference between 
L4 and L5 objects, visible in  Figure~\ref{tColorHist}, 
disappears.} 
\label{FigIncColorHistL4L5}
\end{figure*}

Due to observational selection effects, the L5 subsample of known Trojans
has a larger fraction of objects with large inclinations than the L4
subsample. This difference between L4 and L5, together with the 
color-inclination correlation, results in differences between their  
$t^*$ color distributions discernible in Figure~\ref{tColorHist}. 
However, as shown in Figure~\ref{FigIncColorHistL4L5}, once the objects 
are separated by inclination, or by $\beta$, this difference between 
L4 and L5 objects disappears. We conclude that there is no evidence
for different color-inclination correlations 
between the two swarms.

\begin{figure*} 
\centering
\includegraphics[bb=50 50 401 328, width=8cm]{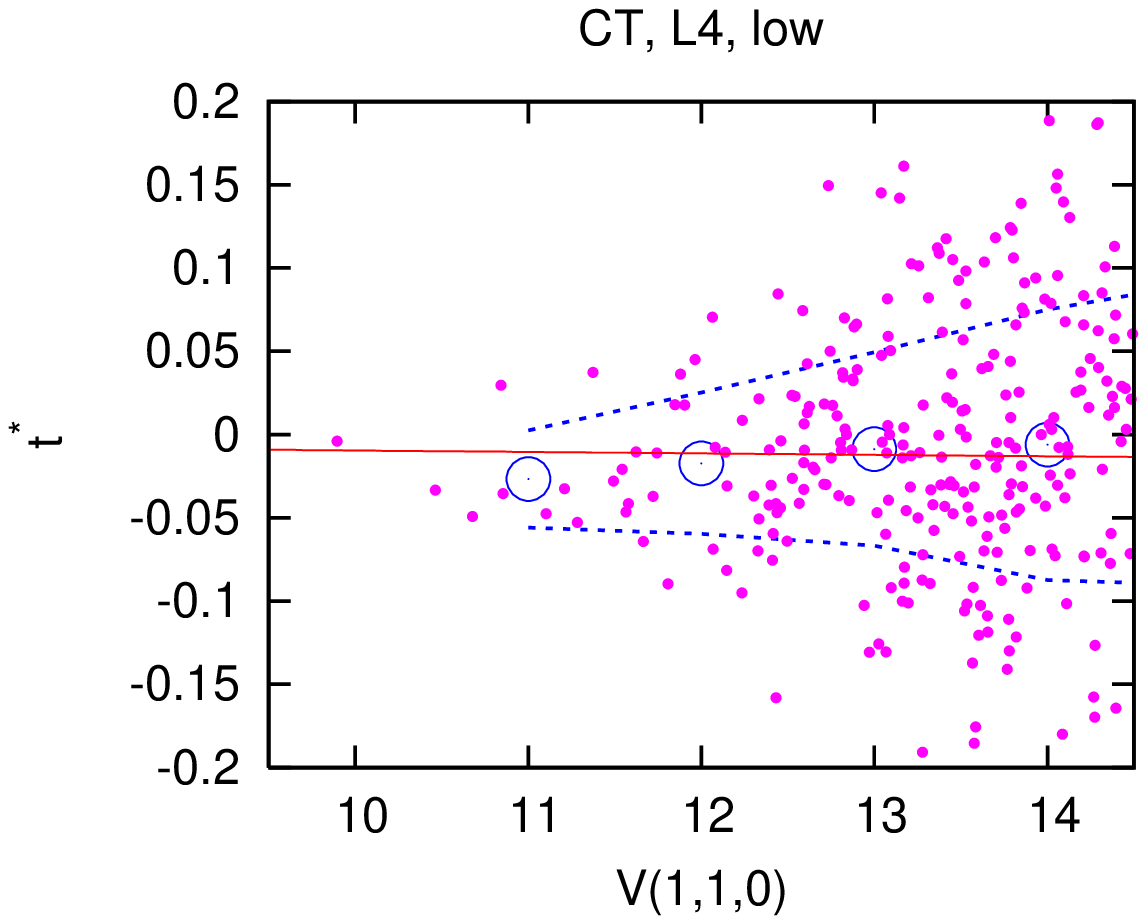}
\includegraphics[bb=50 50 401 328, width=8cm]{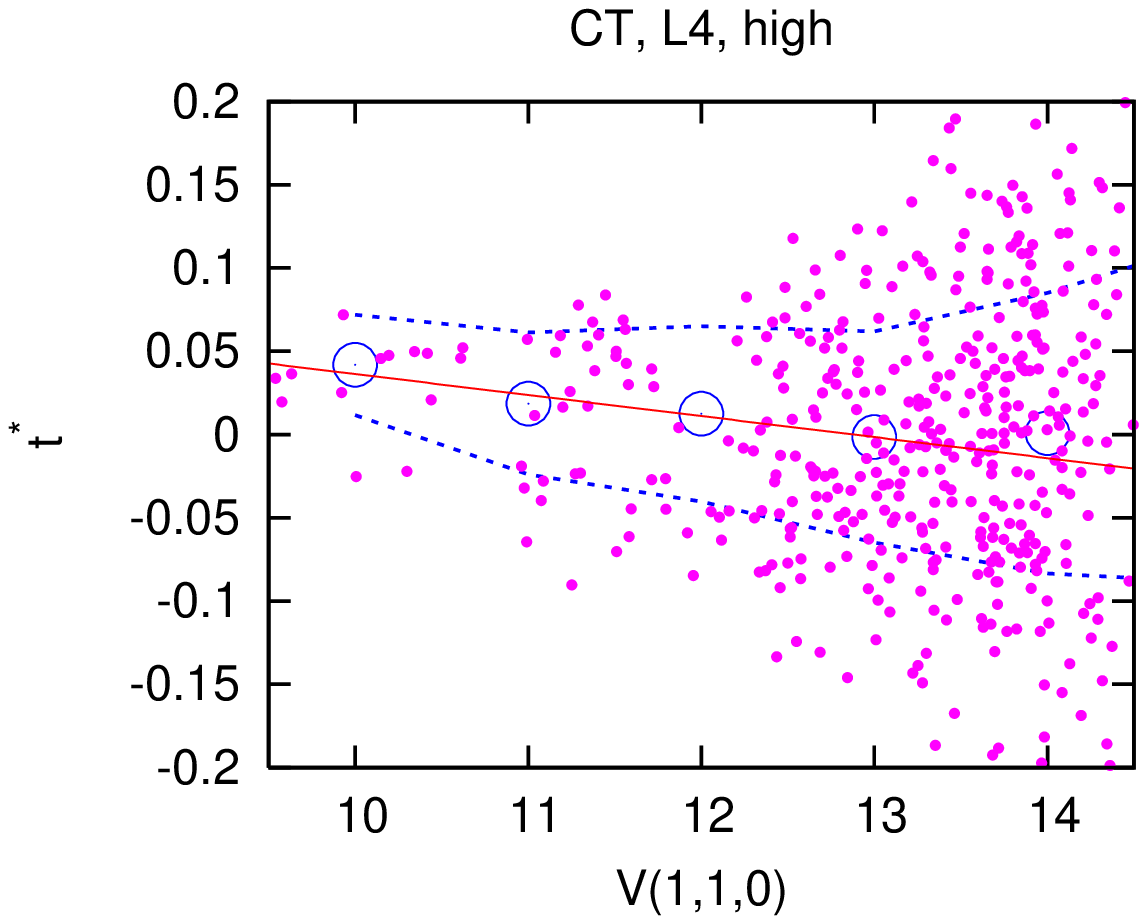}
\includegraphics[bb=50 50 401 328, width=8cm]{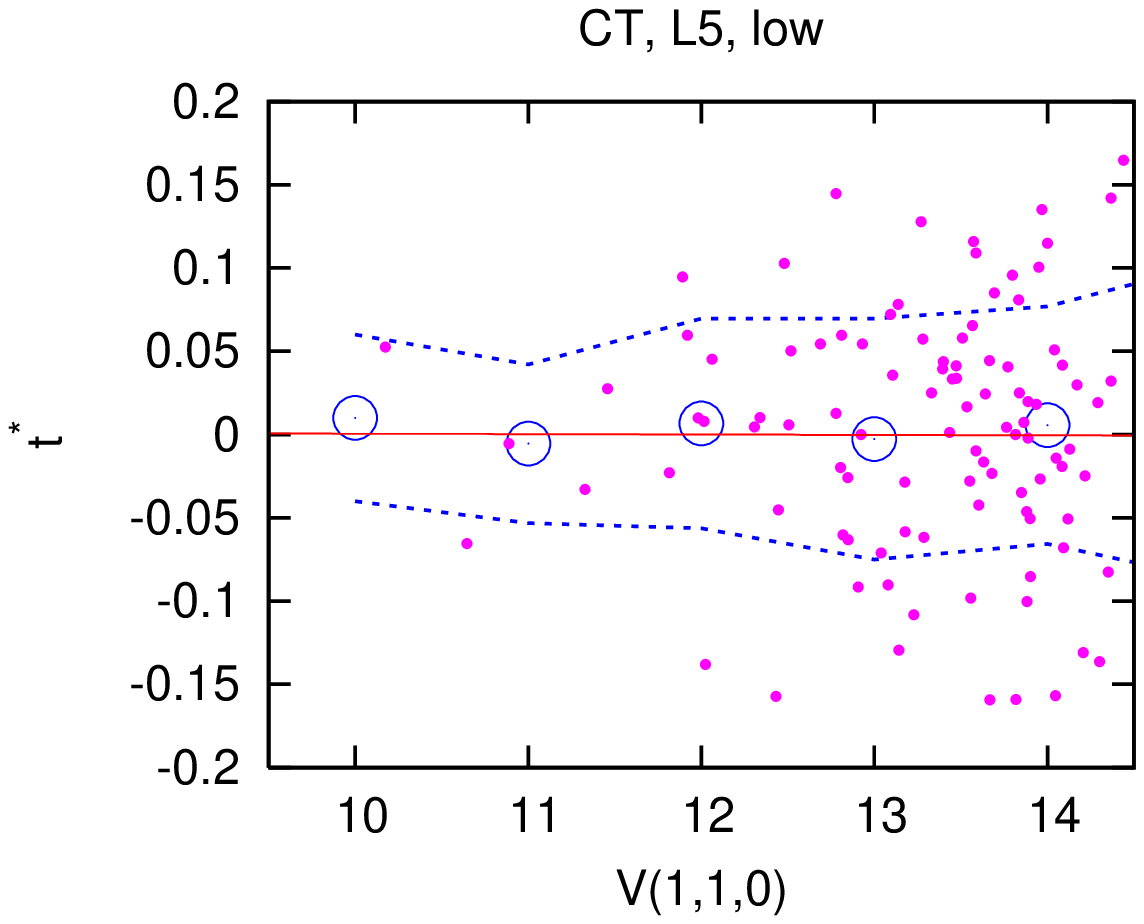}
\includegraphics[bb=50 50 401 328, width=8cm]{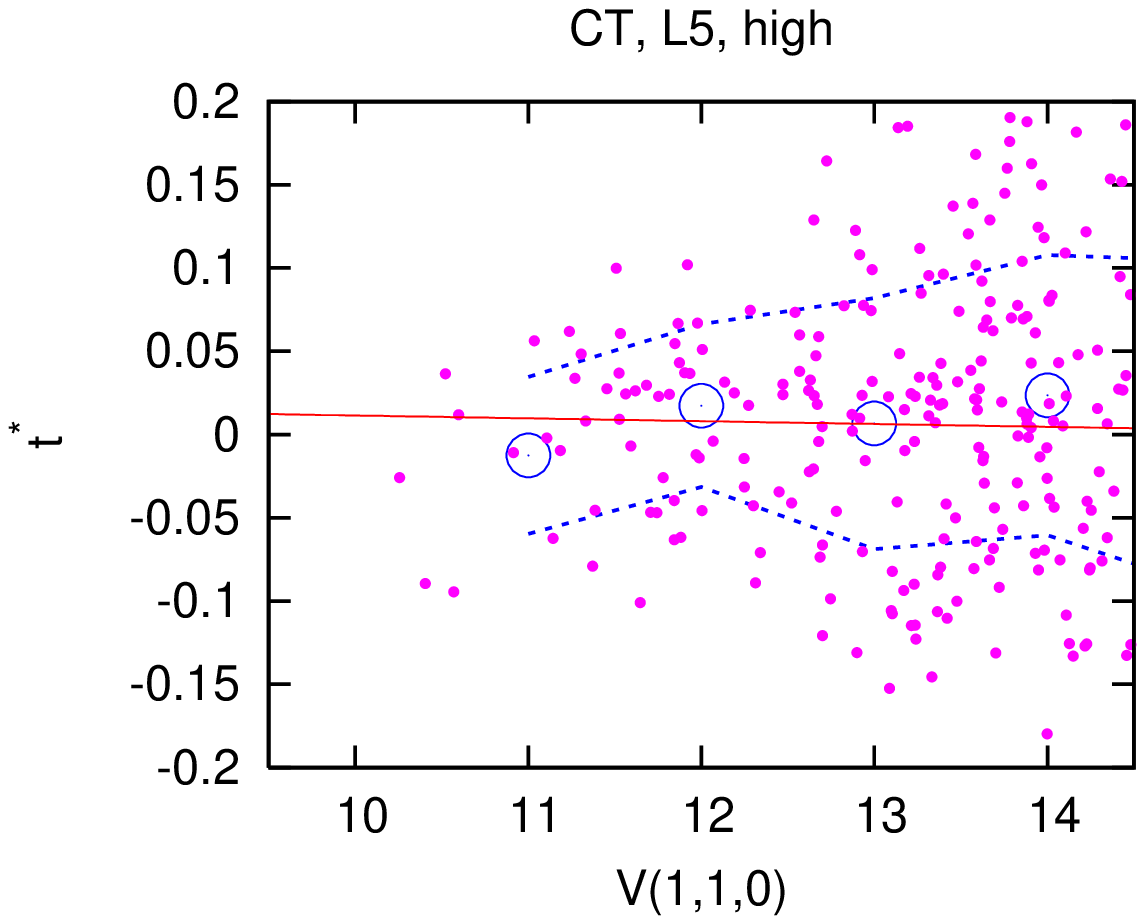}
\caption{Color-magnitude diagrams for subsamples of candidate
Trojans separated into L4 (top) and L5 (bottom) objects, 
and further into low-inclination (left) and high-inclination
(right) objects. Small dots represent individual objects and
large circles are the median values of $t^*$ color in 1 mag
wide bins of absolute magnitude. The 1$\sigma$ envelope around
the median values is computed from the interquartile range. 
Note the cluster of $V(1,1,0)<11$ objects in top right panel
that have slightly redder objects than the rest of the sample.} 
\label{SizeColorFig}
\end{figure*}

The similarity of the histograms shown in the middle and
right panels in Figure~\ref{FigIncColorHist} suggests that
the color-inclination correlation cannot be a strong function
of object's size. Another ``slice'' through the observed
color-inclination-size-swarm space is shown in Figure~\ref{SizeColorFig}.
We find no strong correlation between the Trojan size and color, 
except for a few large L4 objects with high inclination that
have about $\sim$0.05 mag redder $t^*$ color. Indeed, these 
few objects may be the reason for a claim by Bendjoya et al. 
(2004) that the spectral slope (i.e. color) is correlated with 
size in the size range 70--160 km.

\section{       Discussion  and Conclusions                 }

The kinematically-selected sample of candidate Jovian Trojan asteroids
analyzed here is complete at the faint end to $r=21.2$ mag, approximately 
corresponding to 10 km diameter, with a contamination rate of only $\sim$3\%.
Similarity of the longitude (relative to Jupiter) and color distributions 
between known and candidate Trojans,  and their difference from the
distributions for main-belt asteroids which dominate the parent sample, 
strongly suggest that the kinematic selection is robust. The well-controlled 
selection effects, the sample size, depth and accurate five-band UV-IR
photometry enabled several new findings and the placement of 
older results on a firmer statistical footing. The main results obtained
here are:
\begin{enumerate}
 \item{} 
 The differential size distribution of Jovian Trojan asteroids follows a
 power law, $n(D)\propto D^{-q}$, with the power-law index of 
 $q=3.20\pm0.25$, in agreement with previous work (e.g. JTL). This value
 of $q$ implies that the total mass is dominated by large objects. 
 The overall normalization is tied to a complete sample of known Trojans
 and suggests that there are about as many Jovians Trojans as there are 
 main-belt asteroids down to the same size limit, also in agreement 
 with earlier estimates. 

 \item{} 
 The same power-law size distribution provides a good description
 for both the leading (L4) and trailing (L5) swarm. Their spatial 
 distribution on the sky can be described by two elliptical
 Gaussian distributions ($\sigma_\lambda=14^\circ$, $\sigma_\beta=9^\circ$)
 that have {\it different} normalization: {\it there are 
 1.6$\pm$0.1 more objects in the leading than in the trailing swarm.}
The cumulative number of Jovian Trojan asteroids (per deg$^2$) as a 
function of absolute magnitude $H$ and a position in Jupiter's coordinate 
system ($\lambda_J$ and $\beta_J$, in degree) can be estimated from 
 \begin{equation} 
n(H, \lambda_J, \beta_J) = N_{cum}(H) {f(\lambda_J)  \over 2 \pi \sigma_\lambda
      \sigma_\beta } \, \rm{e}^{-{\beta_J^2 \over 2 \sigma_\beta^2}} 
\end{equation} 
where $N_{cum}(H)$ is given for $H<13.5$ by eq.~\ref{Ncum}, and
 \begin{equation} 
f(\lambda_J) = 0.62\,{\rm e}^{-{(\lambda_J-60^\circ)^2 \over 2 \sigma_\lambda^2}}
  + 0.38\,{\rm e}^{-{(\lambda_J+60^\circ)^2 \over 2 \sigma_\lambda^2}}.
\end{equation}
 \item{}
 The two orders of magnitude increase in the number of objects with 
 accurate color measurements allowed us to demonstrate that Trojan asteroids 
 have a remarkably narrow color distribution (root-mean-scatter of only 
 $\sim$0.05 mag) that is significantly different from the color distribution 
 of the main-belt asteroids. 

 \item{}
 We find that the color of Trojan asteroids is correlated with their orbital 
 inclination, in a similar way for both swarms, but appears uncorrelated with 
 the object's size. 
 
 \item{} We did not detect a size-inclination correlation.

\end{enumerate}

These results have direct implications for the theories of 
Trojans origin. The detected difference in the normalization between 
leading and trailing swarms suggests that there was at least 
some period during which their formation and/or evolution was different.
Similarly, the color-inclination correlation suggests that there
must have been a process in the past which is responsible for the
increased fraction of red objects at high orbital inclinations.
Gas dynamics and planetary migration are good candidates for such a 
process, as recently discussed by Tsiganis et al. (2005). A possible 
explanation for this correlation is that when asteroids on the temporary 
eccentric orbits encounter the Sun, their minimal distance from the Sun 
is related to the inclination we observe today. In this picture the space weathering 
effects and volatization would vary with the inclination. A detailed analysis 
of these possibilities is beyond the scope of this paper and we leave it for 
future work. 

While the increase in sample size enabled by SDSS is considerable,
very soon new large-scale sky surveys,  such as Pan-STARRS (Kaiser et
al. 2002) and LSST (Tyson 2002), may obtain even more impressive
samples, both in size, diversity of measurements and their accuracy. 
For example, LSST will scan the whole observable sky every three nights 
in two bands to a $5\sigma$ depth equivalent to $V=25$ (about 2.5 mag
deeper than SDSS). Using the size 
distribution determined here, we estimate that LSST, which may have its 
first light in 2014, will collect a sample of about 100,000 Jovian Trojan 
asteroids and provide {\it both} orbits, accurate color measurements and
light curves for the majority of them. A significant fraction (20--30\%) of 
this sample will be obtained by Pan-STARRS4, which is supposed to have its 
first light around 2009. These samples will undoubtely reinvigorate both 
observational and theoretical studies of Jovian Trojan asteroids.

\section*{Acknowledgments}

We thank Elisabetta Dotto for a discussion that helped improve the 
presentation. This work has been supported by the Hungarian OTKA Grants
T042509, the ``Magyary Zolt\'an'' Higher Educational Public Foundation 
and the Szeged Observatory Foundation. We acknowledge generous support by 
Princeton University.

Funding for the SDSS and SDSS-II has been provided by the Alfred P. Sloan 
Foundation, the Participating Institutions, the National Science Foundation, 
the U.S. Department of Energy, the National Aeronautics and Space 
Administration, the Japanese Monbukagakusho, the Max Planck Society, and 
the Higher Education Funding Council for England. The SDSS Web Site is 
http://www.sdss.org/.

The SDSS is managed by the Astrophysical Research Consortium for the
Participating Institutions. The Participating Institutions are the American
Museum of Natural History, Astrophysical Institute Potsdam, University of
Basel, University of Cambridge, Case Western Reserve University, University of
Chicago, Drexel University, Fermilab, the Institute for Advanced Study, the
Japan Participation Group, Johns Hopkins University, the Joint Institute for
Nuclear Astrophysics, the Kavli Institute for Particle Astrophysics and
Cosmology, the Korean Scientist Group, the Chinese Academy of Sciences
(LAMOST), Los Alamos National Laboratory, the Max-Planck-Institute for
Astronomy (MPIA), the Max-Planck-Institute for Astrophysics (MPA), New Mexico
State University, Ohio State University, University of Pittsburgh, University
of Portsmouth, Princeton University, the United States Naval
Observatory, and the University of Washington.

\end{document}